\documentclass[%
aip,
amsmath,amssymb,
reprint,%
]{revtex4-1}

\usepackage{graphicx}
\usepackage{dcolumn}
\usepackage{bm}

\usepackage[utf8]{inputenc}
\usepackage[T1]{fontenc}
\usepackage{mathptmx}
\usepackage{etoolbox}

\usepackage{braket}
\usepackage[caption=false]{subfig}
\usepackage{dsfont}
\usepackage[table]{xcolor}
\usepackage{placeins}
\usepackage{todonotes}

\newcommand{\refFig}[1]{Fig. \ref{#1}}
\newcommand{\ca}{$c_{{A}}$}
\newcommand{\cb}{$c_{{B}}$}

\newcommand{\req}[1]{Eq. (\ref{#1})}
\newcommand{\refig}[1]{Fig. \ref{#1}}

\makeatletter
\def\@email#1#2{%
	\endgroup
	\patchcmd{\titleblock@produce}
	{\frontmatter@RRAPformat}
	{\frontmatter@RRAPformat{\produce@RRAP{*#1\href{mailto:#2}{#2}}}\frontmatter@RRAPformat}
	{}{}
}%
\makeatother
\begin{document}
	
	\preprint{AIP/123-QED}
	
	\title[Nonequilibrium fluctuations of chemical reaction networks at criticality]{Nonequilibrium fluctuations of chemical reaction networks at criticality: The Schlögl model as paradigmatic case}
	\author{Benedikt Remlein}
	\email{remlein@theo2.physik.uni-stuttgart.de}
	\author{Udo Seifert}%
	\affiliation{ 
		II. Institut für Theoretische Physik, Universität Stuttgart, 70550 Stuttgart
	}%
	\homepage{https://www.itp2.uni-stuttgart.de/}
%
	
	\date{\today}
	
	\begin{abstract}

Chemical reaction networks can undergo nonequilibrium phase transitions upon variation of external control parameters like the chemical potential of a species. We investigate the flux in the associated chemostats that is proportional to the entropy production and its critical fluctuations within the Schlögl model.  Numerical simulations show that the corresponding diffusion coefficient diverges at the critical point as a function of system size. In the vicinity of the critical point, the diffusion coefficient follows a scaling form. We develop an analytical approach based on the chemical Langevin equation and van Kampen's system size expansion that yields the corresponding exponents in the monostable regime. In the bistable regime, we rely on a two-state approximation in order to analytically describe the critical behavior.	
\end{abstract}
	
	\maketitle

	\section{Introduction}
	
Chemical reaction networks constitute a major paradigm for nonequilibrium systems. Due to their complexity, they offer a rich variety of non-trivial phenomena, thus, providing the chance to study nonequilibrium effects.  In the macroscopic limit, chemical reaction networks can exhibit nonlinearities and bifurcations leading to phase transitions that occur out of equilibrium. \cite{schl72,math75,schn76,nico78}

With the development of stochastic thermodynamics \cite{seif12} over the last decades, chemical reaction networks and their thermodynamic treatment gained significant interest.  \cite{bear04,qian05,schm06,rao16,espo20} Several case studies of thermodynamically consistent reaction networks have shown that a phase transition has immediate impact on nonequilibrium quantities such as the coherence of oscillations \cite{nguy18,frit20} or the entropy production rate \cite{croc05,vell09,andr10,rao11,tome12,zhan16,shim16,noa19,herp19,rana20,goes20,dani21,fior21} and its fluctuations. \cite{nguy18,nguy20} From a mathematical perspective, large deviation theory turned out to be a useful tool to examine these phenomena. \cite{ge10a,laza19,jack20,frei21,fior21,guis23} As nonequilibrium phase transitions are not restricted to chemical systems, they can also be found in various biological \cite{heus06,broe11,broe14,bi15,gnes19} and active systems. \cite{butt13,witt14,hube18,gomp20,gopa21,mark21,fodo21a,spec22}

A common feature of chemical reaction networks exhibiting a phase transition is the presence of auto-catalytic reactions and chemostats.  The latter provide a constant source of educt particles to the reaction channels, which, thus, are constantly consuming free energy. The flux and difference in chemical potentials of the respective particles are thus linked to the entropy production of the total system. \cite{seif12,rao16}

So far, the focus has been on the mean behavior of physical observables like entropy production. Only a few studies have examined their fluctuations \cite{nguy20,fior21} for example during  a first order phase-transition. \cite{ge09,iwat10} Fluctuations of the entropy production rate at a continuous phase transition have only been considered numerically in the context of biochemical oscillators. \cite{nguy18} Here, we examine nonequilibrium fluctuations during a continuous phase transition numerically as well as analytically.

We study the flux of particles consumed and produced in the reservoirs for the Schlögl model as a generic univariate system undergoing a continuous phase transition upon variation of a control parameter. Fluctuations of the fluxes are characterized by the associated diffusion coefficient. We numerically find that the particle flux is a continuous function of the control parameter. The diffusion coefficient displays critical behavior while varying a control parameter. We develop a theory that explains the continuous behavior of the particle flux and predicts scaling exponents for the diffusion coefficient in agreement with the numerically found ones.

This paper is organized as follows.  We introduce the model and discuss the different regimes in Sec. \ref{sec:SchlöglModel}. In Sec. \ref{sec:SchlöglNumerics}, we numerically determine the particle flux and the critical exponents for the scaling of the derivative of the flux in the monostable regime. Furthermore, we show how the diffusion coefficient scales with the control parameter and system size. The analytical prediction of scaling exponents for this pathway is presented in Sec. \ref{sec:SchloeglRate} and \ref{sec:SchlöglTheory}. For the transition into bistability,  we numerically show that the flux depends continuously on the control parameters and that the associated diffusion coefficient of the reservoir particles also exhibits critical behavior in Sec. \ref{sec:BiNumerics}. In Sec. \ref{sec:BiMono} and \ref{sec:BiBi}, we analytically describe the behavior of the particle flux and derive the critical exponents of the associated diffusion coefficient for the pitchfork bifurcation into coexistence. In Sec. \ref{sec:scalingform}, we transform into adopted coordinates and derive scaling forms for the particle flux and the diffusion coefficient. We conclude in Sec. \ref{sec:Conclusion}.

\section{Schlögl model}
\label{sec:SchlöglModel}

The reaction scheme which was first proposed by Schlögl \cite{schl72,vell09} consists of an autocatalytic reaction and a linear one,
\begin{equation}
	A + 2X\overset{k_1^+}{\underset{k_1^-}{\rightleftharpoons}} 3X~,~~~~~~X	\overset{k_2^-}{\underset{k_2^+}{\rightleftharpoons}} B~,
	\label{eq:SchloeglCRN}
\end{equation}
for species $X$ in contact with an external reservoir of molecules $A$ and $B$ with fixed concentrations \ca, respectively \cb. The reactions take place in a volume $\Omega$. Due to a difference in the chemical potential between $A$ and $B$ the system is out of equilibrium, i.e., $\Delta \mu \equiv \mu_A - \mu_B > 0$. Considering the cycle in which one $A$ is transformed to a $B$, the thermodynamic driving satisfies the local detailed balance condition\cite{seif12}
\begin{equation}
	e^{\beta \Delta \mu}= \frac{c_A k_1 ^+ k_2^-}{c_B k_1 ^- k_2^+}~,
	\label{eq:LDB}
\end{equation}
with inverse temperature $\beta$ and molecular reaction rates $k_i^\pm$. We set $\beta = 1$ throughout the paper, thus, measuring all energies in units of the thermal energy and entropy in the units of the Boltzmann constant.

We examine the number of produced respectively consumed molecules $A$ and $B$ in the reservoir by introducing random variables $Z_A(t)$ and $Z_B(t)$. Whenever a molecule $A$ is produced, $Z_A(t)$ is decreased by one, whereas, when a molecule $B$ is produced, $Z_B(t)$ is increased by one leading to the following augmented reaction scheme
\begin{equation}
	\{	A + 2X,Z_A-1\} \overset{k_1^+}{\underset{k_1^-}{\rightleftharpoons}} \{3X,Z_A\}~,~\{X,Z_B\}	\overset{k_2^-}{\underset{k_2^+}{\rightleftharpoons}} \{B,Z_B+1\}~.
	\label{eq:SchloeglAB}
\end{equation} 
The mean flux entering and leaving the particle reservoirs is given by
\begin{equation}
	J_B \equiv  \lim\limits_{t\to \infty} \frac{\braket{Z_B(t)}}{\Omega t} = J_A~,
	\label{eq:ZJB}
\end{equation}
with fluctuations quantified by the diffusion coefficient
\begin{equation}
	D_B \equiv \lim\limits_{t\to \infty} \frac{\mathrm{Var}[Z_B(t)]}{2\Omega t} = D_A~,
	\label{eq:ZDB}
\end{equation}
where $\mathrm{Var}[z] \equiv \braket{z^2} - \braket{z}^2$. The equality between $D_A$ and $D_B$ has been proven in Ref. \cite{nguy20} using large deviation theory. Since the particle fluxes and their fluctuations are equal in both reservoirs, we consider in the rest of the paper only particles of species $B$. All results are equally true for the particle flux of species $A$. Thus, we drop the index $B$ for the mean values in the following.

The entropy production 
\begin{equation}
	\Delta S(t) = Z_A(t)\mu_A - Z_B(t)\mu_B
\end{equation}
is proportional to the particle flux, thus, it exhibits the same fluctuations. The mean entropy production rate in the steady state is determined as
\begin{equation}
	\sigma \equiv \lim\limits_{t\to\infty} \frac{1}{t}\braket{\Delta S(t)} =  \Omega J \Delta \mu~.
	\label{eq:EPR}
\end{equation}

In the thermodynamic limit, i.e., for $\Omega \to \infty$, the reaction scheme, \req{eq:SchloeglCRN}, is described by the deterministic rate equation
\begin{equation}
	\partial_t \hat x(t) = -k_1^- \hat x(t)^3 + c_A k_1^+ \hat x(t)^2 - k_2^-\hat x(t) + c_Bk_2^+,
	\label{eq:SchloeglRateEquation}
\end{equation}
where $\hat x(t)$ denotes the mean concentration of species $X$ at time $t$. We set $k_i^\pm = 1$ throughout the paper which effectively corresponds to a rescaling of the parameters as
\begin{eqnarray}
	\tilde c_A &\equiv& c_A k_1^+/(k_1^- k_2^-)^{\frac 12}~,~~~~~~~~
	\tilde c_B \equiv c_B (k_1^-)^{\frac 12}k_2^+/(k_2^-)^{3/2}~, \nonumber\\
	\tilde \Omega &\equiv& \Omega {(k_2^-)^{\frac 12}/(k_1^-)^{\frac 12}}~, ~~\text{and}~~
	\tilde t \equiv t k_2^-~.
\end{eqnarray}
Thus, the rate equation (\ref{eq:SchloeglRateEquation}) depends only on the rescaled concentrations $\tilde c_A$ and $\tilde c_B$. We drop the tilde in the following.

\begin{figure}
	\centering
	\includegraphics[width=0.45\textwidth]{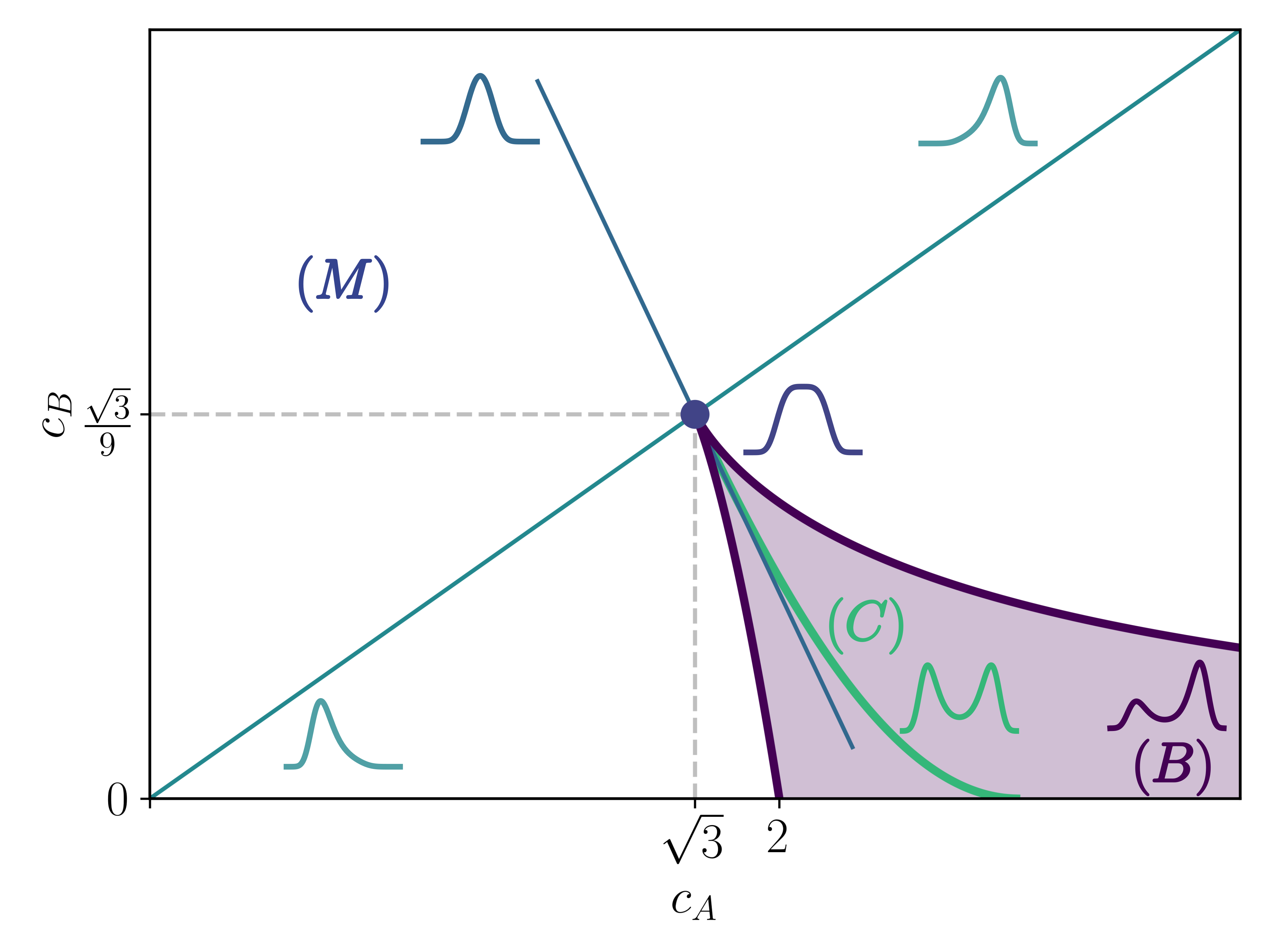}
	\caption{Phase diagram for the Schlögl model, \req{eq:SchloeglCRN}. The monostable regime is denotes as $(M)$. The colored area is the bistable regime, $(B)$. $(C)$ labels the coexistence line. Where the phase boundaries of the bistable region meet, the system exhibits a critical point (blue dot). The green line through the origin and this point is denoted critical line. The blue line is the tangential line into the coexistence area. For the specific areas, the typical shape of the steady-state probability distribution $p^s(x)$ is sketched.}
	\label{fig:pd}
\end{figure}

Depending on the choice of concentrations, \req{eq:SchloeglRateEquation} shows a different number of steady-states, see \refig{fig:pd}. The system exhibits a monostable phase $(M)$ with a unique, stable fixed point of the dynamics, \req{eq:SchloeglRateEquation}. The second phase is the cusp-like region with two stable fixed points and an unstable one in between $(B)$. See Appendix \ref{app:pd} for a derivation of the phase boundaries. Within the bistable region, there is a line denoted as coexistence line $(C)$ along which the two stable fixed points correspond to the global minimum of the dynamics. Away from this line there exists only one global minimum. At the end of the cusp-like region where the coexistence line ends, there is a critical point with coordinates 
\begin{equation}
c_A^{\mathrm{crit}} \equiv \sqrt{3}\text{~~~and~~~}c_B^{\mathrm{crit}} \equiv {\sqrt{3}}/{9}~.
\end{equation}

From a thermodynamic perspective, there occur different phase transitions depending on the path chosen within the $c_A-c_B$-plane. Crossing the coexistence line, the system undergoes a first order phase transition. While this transition has been the topic of several studies, \cite{ge09,vell09,nguy20} we here consider two pathways through the critical point, i.e., the continuous phase transition. The first path corresponds to a constant thermodynamic driving
\begin{equation}
	\Delta \mu_{\mathrm{crit}} \equiv \ln 9~,
\end{equation}
to which we refer as the critical line.

As second path, we consider the line which enters the cusp-like region tangential to its boundary and goes through the critical point, see the blue line in \refig{fig:pd}. Along that line, the system undergoes a pitchfork bifurcation. 

\section{Along the critical line}

We consider the line in the $c_A-c_B$-plane parameterized as
\begin{equation}
	c_B(c_A) \equiv c_A/9~,
	\label{eq:CritLine}
\end{equation}
along which the concentration \ca~and system-size $\Omega$ remain as control parameters.

\subsection{Numerical data}
\label{sec:SchlöglNumerics}
\begin{figure*}
	\centering
	\subfloat[\label{sfig:1a}]{%
		\includegraphics[width=.325\textwidth]{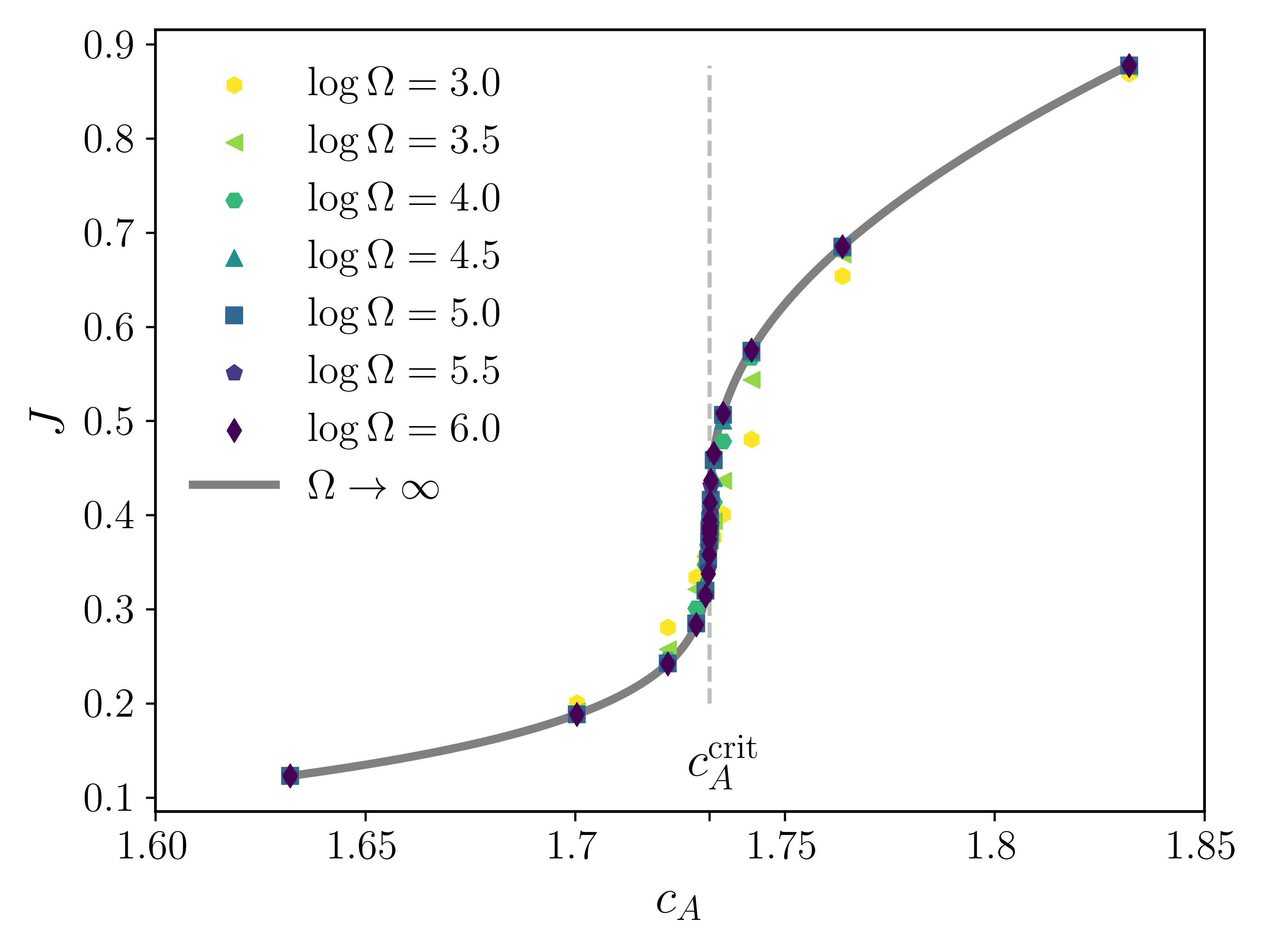}
	}
	\subfloat[\label{sfig:1b}]{%
		\includegraphics[width=.325\textwidth]{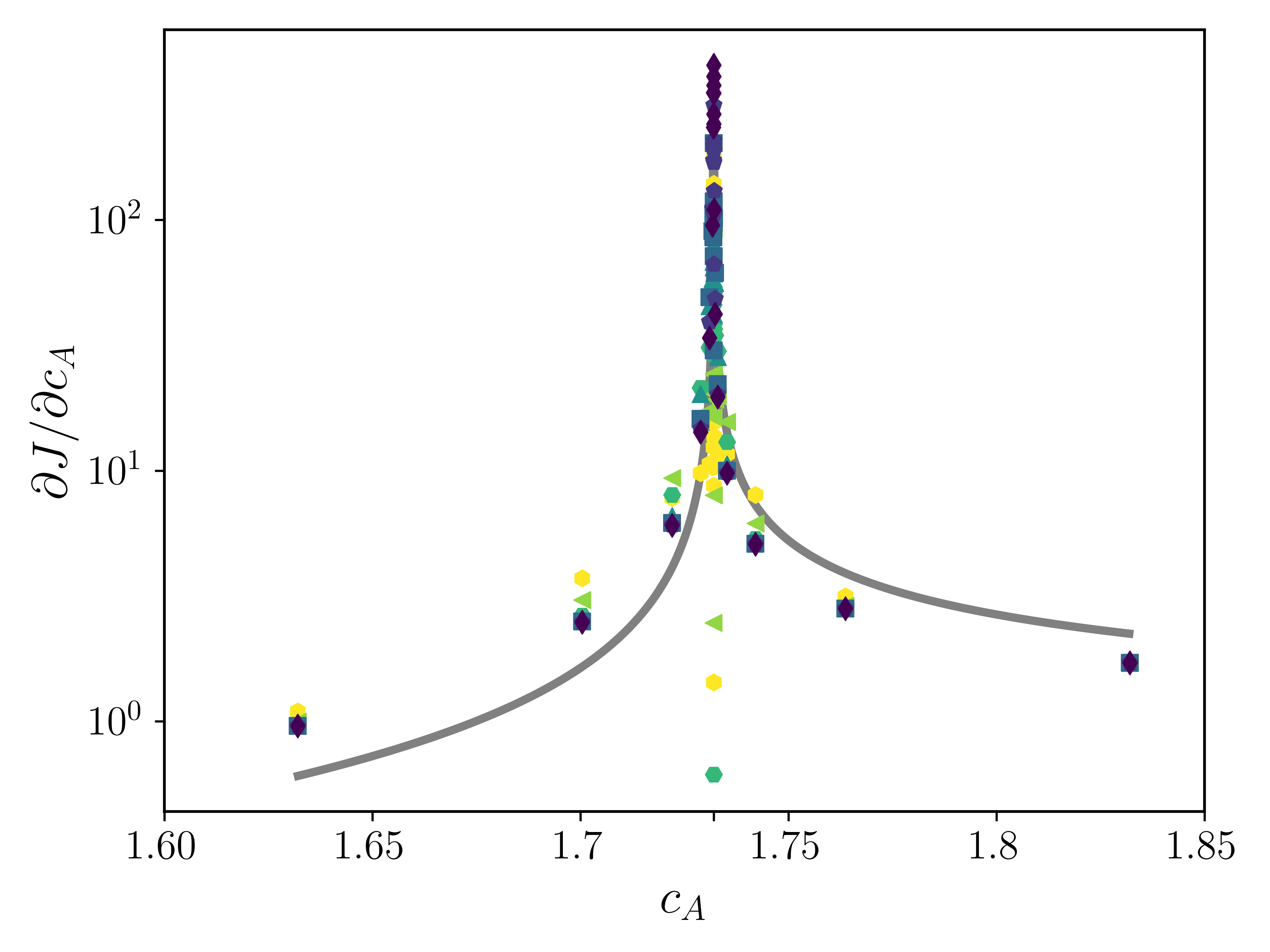}
	}
	\subfloat[\label{sfig:1c}]{%
		\includegraphics[width=.325\textwidth]{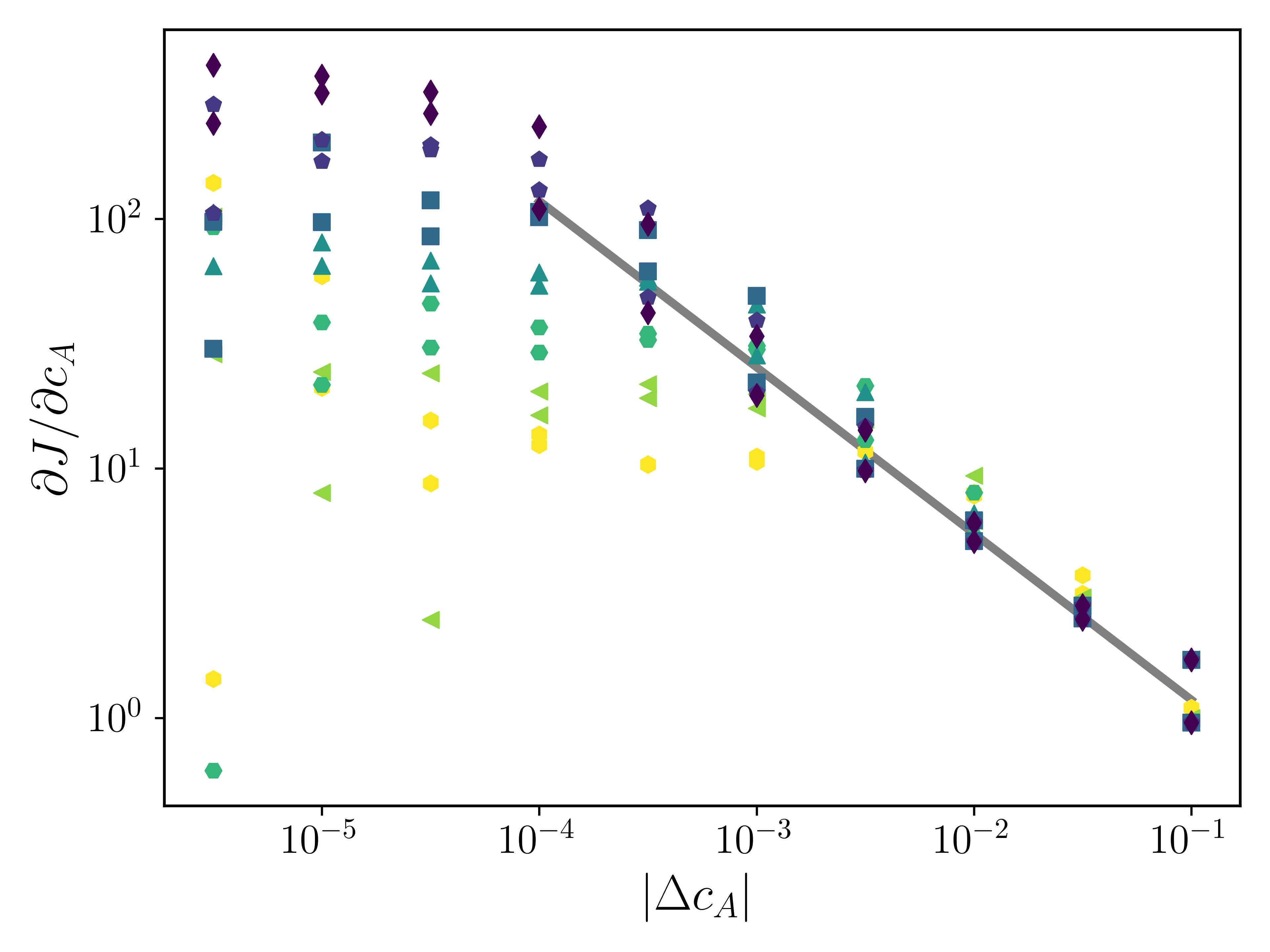}}
	\caption{Mean flux $J$ of the Schlögl model along the critical line. (a) $J$ as function of the control parameter $c_A$ for various system sizes $\Omega$. (b) Numerical derivative of $J$ with respect to the control parameter $c_A$.  (c) Derivative of $J$ as function of  $\Delta c_A$. The concentration $ c_B$ is determined according to the local detailed balance condition, Eq. (\ref{eq:LDB}). A solid line represents the analytical prediction, \req{eq:CritJB}, respectively \req{eq:CritDJB}. }
	\label{fig:1}
\end{figure*}
\begin{figure}
	\centering
	\subfloat[\label{sfig:2a}]{%
		\includegraphics[width=.4\textwidth]{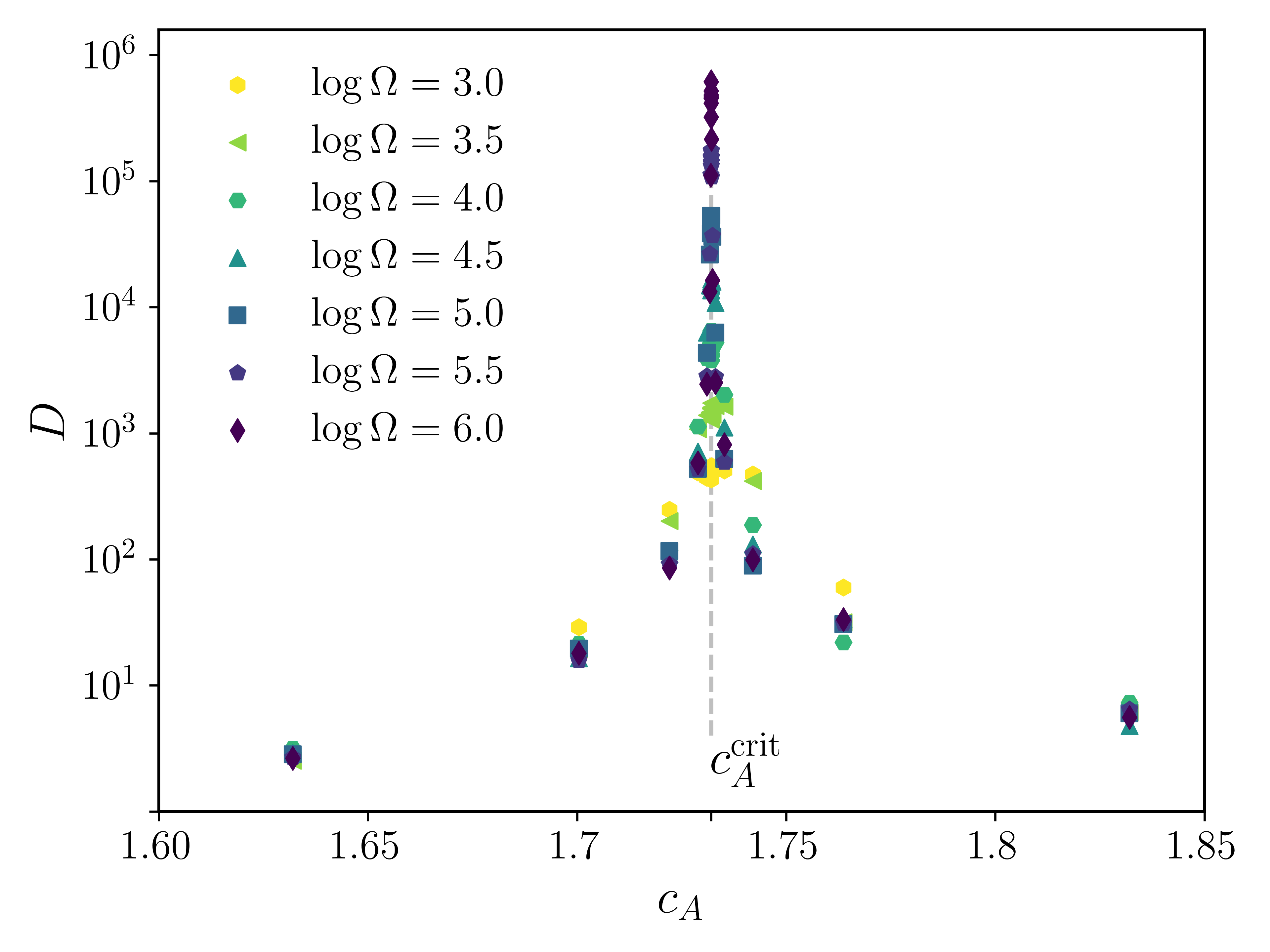}
	}	
	
	\subfloat[\label{sfig:2b}]{%
		\includegraphics[width=.4\textwidth]{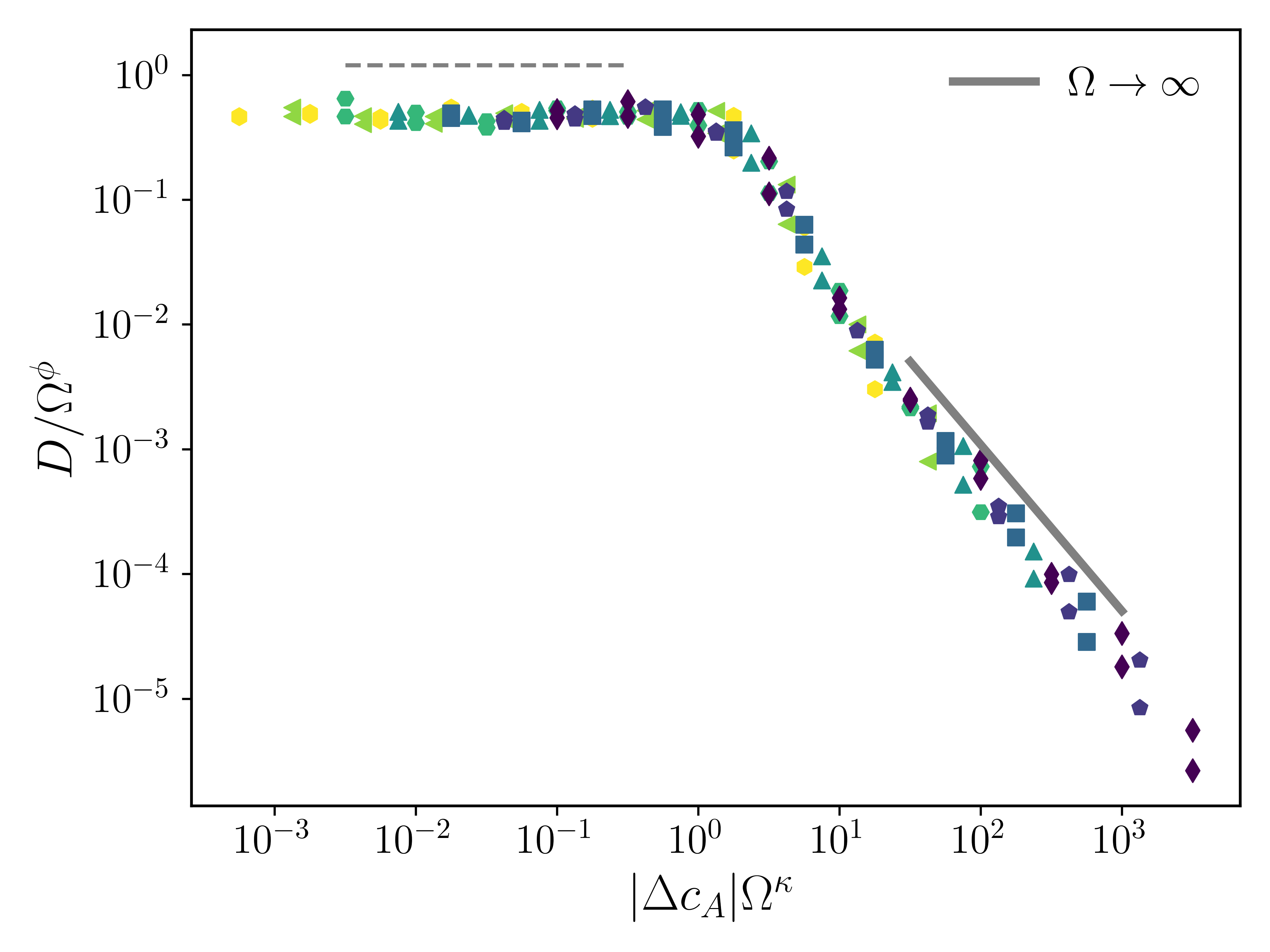}}
	\caption{Diffusion coefficient $D$ of the Schlögl model along the critical line. (a) $D$ as function of the control parameter $c_A$ for various system sizes $\Omega$. (b) Scaling of $D$ according to $ D(\Delta  c_A, \Omega) = \Omega^\phi \mathcal D(|\Delta  c_A|~\Omega^\kappa)$ with $ c_A^{\mathrm{crit}} = \sqrt{3}$, $\kappa = 0.75$ and $\phi = 1.00$. $ c_B$ is determined according to the local detailed balance condition, Eq. (\ref{eq:LDB}). The solid line represents the analytical prediction, \req{eq:SchloeglDBNotCrit}.}
	\label{fig:2}
\end{figure}

	
Using Gillespie's direct method, \cite{gill77} we integrate the reaction scheme Eq. (\ref{eq:SchloeglCRN}) numerically for various values of the control parameters. We determine $Z_B(t)$ according to Eq. (\ref{eq:SchloeglAB}) and calculate the mean flux $J$ and diffusion coefficient $D$ in the steady-state through  \req{eq:ZJB}, respectively (\ref{eq:ZDB}). The results are shown in \refFig{fig:1} and \ref{fig:2}. 

	At the critical point, the mean flux $J$  has a vertical tangential, i.e., a diverging first derivative with respect to the control parameter $c_A$, see \refFig{sfig:1b}. Small deviations of the numerical derivative from the theoretical line are due to the non-uniform step-size while performing the numerical derivative. Shifting the $c_A$-axis in \refFig{sfig:1c}, we extract a power-law for the divergence in $\Delta c_A \equiv  c_A -  c_A^{\mathrm{crit}}$ independent on the reaction volume $\Omega$, i.e.,
	\begin{equation}
		\frac{\partial J}{\partial c_A} \propto |\Delta c_A|^{-\chi}~,~~\chi = 0.65 \pm 0.05~.\label{eq:criticalChi}
	\end{equation}
	We attribute the scattering of the derivative for $|\Delta c_A|\ll 1$ in \refig{sfig:1c} to the noisy numerical derivative on the logarithmic scale.
	
	The diffusion coefficient $D$ displays a pronounced peak around the critical value, see \refFig{sfig:2a}.
	The maximum depends on the reaction volume $\Omega$. Off the critical point, i.e., for $\Delta  c_A \neq 0$, the value of $D$ is  independent of $\Omega$. Furthermore, the data obey a scaling relation of the form
	\begin{equation}
		D = D(\Delta  c_A, \Omega) = \Omega^\phi \mathcal D(|\Delta  c_A|~\Omega^\kappa)~
	\end{equation}
	as the collapse onto a master curve shows, see \refFig{sfig:2b}. We determine the exponents as
	\begin{equation}
		\phi = 1.0 \pm 0.1~~\text{and}~~\kappa = 0.75 \pm 0.05~,
		\label{eq:SchloeglNumericExp}
	\end{equation}
	which results in a scaling function 
	\begin{equation}
		\mathcal D(x) \propto	\begin{cases}
			1~,& |x| \ll 1\\ x^\varepsilon ~,& |x| \gg 1
		\end{cases}~
		\label{eq:SchloeglScalingForm}
	\end{equation}
	with 
	\begin{equation}
		\varepsilon = - \phi/\kappa = -1.3\pm 0.2~.
		\label{eq:SchloeglNumericExp2}
	\end{equation}
	The plateau for small values of $|\Delta c_A|\Omega^\kappa$ indicates a scaling of the critical value as $D^{\mathrm{max}} \propto \Omega^\phi$ with $\phi$ given in Eq. (\ref{eq:SchloeglNumericExp}). The power-law behavior with $-\phi/\kappa$ represents a scaling off the critical point with $D \propto \Omega^0$ consistent with the raw data, \refFig{sfig:2a}. 

	In the next two sections, we approximate the reaction scheme (\ref{eq:SchloeglAB}) through the rate equation and as a diffusion process to determine these exponents analytically.

	\subsection{Rate equation and mean flux}
	\label{sec:SchloeglRate}
	In the limit $\Omega \to \infty$, the mean concentration $\hat x(t)$ obeys the reaction rate equation
	\begin{equation}
		\partial_t \hat x(t) = -[\hat x(t) - 1/\sqrt3]^3 + \Delta  c_A [\hat x(t)^2 + 1/9]~.
		\label{eq:alpha1}
	\end{equation}
	The reaction scheme (\ref{eq:SchloeglCRN}) has a unique fixed point $x^*$, thus, the rate equation approximates the system well in the limit of large reaction volume. 
	Above the critical point, i.e., for $\Delta c_A >0$, the fixed point is
	\begin{eqnarray}
	x^*(\Delta c_A)& =& 
	{\Delta c_A}/{3}+[\Delta c_A^2 (\Delta c_A+2 \sqrt 3)]^{1/3}/3	\label{eq:FPabove}\\ && +
	[\Delta c_A (\Delta c_A^2+4\sqrt 3 \Delta c_A+12)]^{1/3}/3+1/\sqrt 3\nonumber
\end{eqnarray}
	For $-\sqrt{3}\leq \Delta c_A < 0$, we get
	\begin{eqnarray}
		x^*(\Delta c_A) &=& {\Delta c_A}/{3}+{[-\Delta c_A (\Delta c_A+2\sqrt 3)^2]^{2/3}}/{3 (\Delta c_A+2\sqrt 3)}\nonumber\\&& -[-\Delta c_A (\Delta c_A+2\sqrt 3)^2]^{1/3}/3+1/\sqrt 3~.
		\label{eq:FPbelow}
	\end{eqnarray}
	For $|\Delta c_A|\ll 1$, we find
	\begin{eqnarray}
		x^*(\Delta c_A) &=& 1/\sqrt 3 + \mathrm{sign}(\Delta c_A) 2^{2/3} |\Delta c_A|^{1/3} /3^{2/3} 	\label{eq:SchloeglFP}\\ &&+ 2^{1/3}|\Delta c_A|^{2/3}/3^{5/6}  + \mathcal O(|\Delta c_A|)~.\nonumber
	\end{eqnarray}
	Thus, for $\Delta c_A \to 0$, we get a continuous transition from one solution to the other, i.e.,
	\begin{equation}
		\lim\limits_{\Delta c_A\downarrow 0}x^*(\Delta c_A) = 	\lim\limits_{\Delta c_A \uparrow 0}x^*(\Delta c_A) = 1/\sqrt{3}~. 
		\label{eq:FPcont}
	\end{equation}	
	The rate equation for the variable $Z_B$ reads
	\begin{equation}
		\partial_t \hat z_B(t) =  \hat x(t)- c_A/9~,
		\label{eq:ZBRATEEQ}
	\end{equation}
	where $\hat z_B(t) \equiv \braket{Z_B(t)}/\Omega$ denotes the mean density of produced particles $B$ up to time $t$. In the stationary state, we find
	\begin{equation}
	\hat z_B(t) = [ x^*(\Delta c_A)-(\sqrt{3}+\Delta c_A)/9] t~,
	\label{eq:ZBrate}
	\end{equation}
	leading to a mean flux
	\begin{equation}
		J = J(\Delta c_A) \equiv x^*(\Delta c_A)-(\sqrt{3}+\Delta c_A)/9
		\label{eq:CritJB}
	\end{equation}
	as a function of the control parameter. This expression together with Eqs. (\ref{eq:FPabove}-\ref{eq:FPbelow}) gives the theoretical prediction of the mean flux along the critical line shown in \refFig{fig:1}.
	For the derivative of $J$ we obtain the scaling law
	\begin{equation}
		\frac{\partial J}{\partial c_A}(\Delta c_A) =\frac{2^{2/3}}{3^{5/3}}|\Delta c_A|^{-2/3}
		\label{eq:CritDJB}
	\end{equation}
	as displayed in \refFig{sfig:1c}. Thus, we obtain the scaling exponent
	\begin{equation}
	\chi = 2/3~,
	\label{eq:chi}
	\end{equation}
	in agreement with the numerically derived one, \req{eq:criticalChi}.
	
	\subsection{Diffusion approximation and diffusion coefficient}
	\label{sec:SchlöglTheory}
	
	\subsubsection{Set-up}
	We approximate the augmented reaction scheme, Eq. (\ref{eq:SchloeglAB}), by Gillespie's chemical Langevin equation \cite{gill00,horo15} and van Kampen's system size expansion. \cite{vankampen,horo15} The chemical Langevin equation for the number of molecules of species $B$ reads
	\begin{equation}
		\partial_t Z_B(t) = W_2^+[n(t)] - W_2^-[n(t)] + \sqrt{W_2^+[n(t)]+ W_2^-[n(t)]}\xi_B(t)~,
		\label{eq:CLEB}
	\end{equation}
	with Gaussian white noise $\xi_B(t)$, i.e., $\braket{\xi_B(t)} = 0$ and $\braket{\xi_B(t)\xi_B(t^\prime)} = \delta(t-t^\prime)$. The $W_i^\pm[n(t)]$ are the propensity functions defined through Eq. (\ref{eq:SchloeglAB}) with the respective reaction direction, i.e.,
	\begin{eqnarray}
		W_1^+(n) &\equiv& c_A n(n-1)/\Omega~,\nonumber\\
		W_1^-(n) &\equiv& n(n-1)(n-2)/\Omega^2~,\nonumber\\
		W_2^+(n) &\equiv&  \Omega c_B~~~~~~\text{and}\\
		W_2^-(n) &\equiv&n~,\nonumber
	\label{eq:JumpRates}
	\end{eqnarray}
	where we set $c_B = c_A/9$. Here, $n$ denotes the number of molecules of species $X$.
	
	In analogy to van Kampen's system size expansion, \cite{vankampen,horo15} we make the ansatz
	\begin{eqnarray}
		\begin{pmatrix}n(t)\\Z_B(t)\end{pmatrix} &\equiv& \Omega\begin{pmatrix}\hat x(t)\\\hat z_B(t)\end{pmatrix} + \Omega^\mu \begin{pmatrix} x(t)\\z_B(t)\end{pmatrix}
		\label{eq:vanKampenAnsatz}
	\end{eqnarray}
	with a scaling exponent $0 < \mu < 1$. The idea of this approximation is to calculate the fluctuations of an observable around a macroscopic trajectory. Plugging the ansatz, Eq. (\ref{eq:vanKampenAnsatz}), into the chemical Langevin equation (\ref{eq:CLEB}) and dividing by $\Omega$, we obtain after a Taylor expansion for $\Omega \gg 1$ in lowest order the mean field equation, i.e., the $\mathcal O(1)$ for $\Omega \to \infty$ leads to the rate equation (\ref{eq:ZBRATEEQ}).
	 
	The higher order terms can be rearranged as
	\begin{eqnarray}
		\partial_t z_B(t) &=& \{W_2^{+~\prime}[\hat x(t)]-W_2^{-~\prime}[\hat x(t)]\}x(t) \\
		&& + \Omega^{1/2 -\mu} \sqrt{ W_2^{+}[\hat x(t)]+W_2^{-}[\hat x(t)] + \mathcal O(\Omega^{\mu -1})} \xi_B(t)~,\nonumber
	\end{eqnarray}
	where a prime denotes a derivative with respect to the argument. In order to obtain an equation for the fluctuations which is independent on $\Omega$, we set $\mu = 1/2$. This exponent is consistent with van Kampen's linear noise approximation. \cite{vankampen,horo15} Thus, with the dynamical equation above and the definition of $D$, Eq. (\ref{eq:ZDB}), we find
	\begin{widetext}
		\begin{eqnarray}
			\label{eq:FullDB}
			D &= &\frac 12\lim\limits_{t\to \infty}\frac 1t\int\limits_0^t d\tau\int\limits_0^t ds  \bigg( W_2^{+~\prime}[\hat x(\tau)]-W_2^{-~\prime}[\hat x(\tau)]\bigg)\bigg( W_2^{+~\prime}[\hat x(s)]-W_2^{-~\prime}[\hat x(s)]\bigg)~\braket{x(\tau)x(s)} \\
			&&\quad + \frac 12 \lim\limits_{t\to \infty}\frac 1t\int_0^t ds \bigg(W_2^{+}[\hat x(s)]+W_2^{-}[\hat x(s)]\bigg) \nonumber \\ && \quad {-  \lim\limits_{t\to \infty}\frac 1t\int\limits_0^t d\tau\int\limits_0^t ds  \bigg( W_2^{+~\prime}[\hat x(\tau)]-W_2^{-~\prime}[\hat x(\tau)]\bigg)\sqrt{\frac 12 \bigg(W_2^{+}[\hat x(s)]+W_2^{-}[\hat x(s)]\bigg)}~\braket{x(\tau)\xi_B(s)} }  \nonumber~.
		\end{eqnarray}
	\end{widetext}
	
	In order to obtain the diffusion coefficient $D$, we need to examine the dynamics of $\hat x(t)$ and $x(t)$, which is independent on $Z_B(t)$. Let $p(n,t)$ be the probability to have $n$ molecules of species $X$ at time $t$. Using the ansatz, Eq. (\ref{eq:vanKampenAnsatz}), and the transformation rule for probability densities, we find for the density of the fluctuations
	\begin{equation}
		\pi(x,t) = \Omega^{1/2} p(\Omega \hat x + \Omega^{1/2}x,t)~.
	\end{equation}
	Following van Kampen's system-size expansion, we obtain the corresponding Fokker-Planck equation
	\begin{eqnarray}
		\partial_t \pi(x,t) & =& \frac{d}{dt} \Omega^{1/2}  p[\Omega \hat{x}(t) + \Omega^{1/2} x,t]\nonumber\\
		& = & \Omega^{1/2} \partial_x\bigg(\{\partial_t \hat{x}(t) - {\alpha}_{1}[\hat{x}(t)]\} \pi(x,t) \bigg)\nonumber\\
		&& - \partial_x\bigg[ \alpha_{1}^{\prime}[\hat{x}(t)] x~ \pi(x,t) + \mathcal O(\Omega^{-1/2})\bigg]\nonumber\\
		&& + \frac 12 \partial_x^2 \bigg[ \alpha_{2}[\hat{x}(t)]\pi(x,t) + \mathcal O(\Omega^{-1/2})\bigg]\nonumber\\
		&& + \mathcal O(\Omega^{-1/2})~.
		\label{eq:SchloeglXLinNoiseEq}
	\end{eqnarray}
	The jump-moments are defined through
	\begin{equation}
		\lim\limits_{\Omega \to \infty}	\Omega  \alpha_1(n/\Omega) \equiv \int d n^\prime (n - n^\prime)W(n|n^\prime)
	\end{equation}
	and
	\begin{equation}
		\lim\limits_{\Omega \to \infty}\Omega  \alpha_2(n/\Omega) \equiv \int d n^\prime ( n- n^\prime)^2W(n| n^\prime)~
	\end{equation}
	The jump-rates $W(n|n^\prime)$ contain the propensity functions and are defined through Eq. (\ref{eq:SchloeglCRN}) as
	\begin{equation}
		W(n|n^\prime) \equiv \delta_{n^\prime,n-1} [W_1^+(n) + W_2^+(n)] + \delta_{n^\prime,n+1} [W_1^-(n) + W_2^-(n)]~.
	\end{equation}
	$\delta_{i,j}$ denotes the Kronecker-delta. 
	
	The condition for the curly bracket in the second line of Eq. (\ref{eq:SchloeglXLinNoiseEq}) to vanish determines the macroscopic law for the $X$ species, i.e., the rate equation (\ref{eq:alpha1}). As discussed above, this equation has a stable fixed point $x^*$, i.e., $\hat x(t) \to x^*$ for $t \to \infty$. Thus, the motion of $z_B(t)$ is in the long time limit equivalent to the following Langevin equation
	\begin{equation}
	\partial_t z_B(t) = 		 ( W_2^{+~\prime}[x^*]-W_2^{-~\prime}[x^*]) x(t) + \sqrt{  W_2^{+}[x^*]+W_2^{-}[x^*]}\xi_B(t)~.
		\label{eq:SchloeglLangevin}
	\end{equation}
	The diffusion coefficient, Eq. (\ref{eq:FullDB}), reduces to
	\begin{eqnarray}
		D &=& ( W_2^{+~\prime}[x^*]-W_2^{-~\prime}[x^*])^2 \int\limits_0^\infty dt\braket{x(t)x(0)} \nonumber\\ 
		&& + \frac12  ( W_2^{+}[x^*]+W_2^{-}[x^*]) 	\label{eq:SchloeglDB}\\
		&& - \sqrt{\frac 12  ( W_2^{+}[x^*]+W_2^{-}[x^*])} ( W_2^{+~\prime}[x^*]-W_2^{-~\prime}[x^*]) \nonumber \\ && \quad \times \lim\limits_{t\to \infty}\frac 1t\int\limits_0^t d\tau\int\limits_0^t ds~\braket{x(\tau)\xi_B(s)}  \nonumber~.	
	\end{eqnarray}
	
	In the next paragraphs, we will examine the fluctuations $x(t)$ around the fixed point $x^*$ both for $c_A =  c_A^{\mathrm{crit}}$, and for $c_A \neq c_A^{\mathrm{crit}}$ in order to obtain an expression for the integral over the correlation function $\braket{x(t)x(0)}$ and $\braket{x(\tau)\xi_B(s)}$.
	
	\subsubsection{At the critical point}
	
	For $c_A = c_A^{\mathrm{crit}} = \sqrt{3}$, the mean-field equation, Eq. (\ref{eq:alpha1}), reduces to
	\begin{equation}
	\partial_t\hat x(t) = \alpha_1[\hat x(t)] = -[\hat x(t)-1/\sqrt 3]^3~.
		\label{eq:SchlöglCrit2}
	\end{equation}
	We assume that the system has already relaxed into the stationary state of reaction scheme (\ref{eq:SchloeglCRN}). This state corresponds to the fixed point $x^* \equiv 1/\sqrt 3$. Thus, we can replace $\hat x(t) \to x^*$. The first and second derivative of the first jump moment vanishes at the critical point, compare \req{eq:SchlöglCrit2}. Following the strategy outlined above, we obtain for the probability density of the fluctuations
	\begin{eqnarray}
		\partial_t \pi(x,t) &=& | \alpha_1^{\prime\prime\prime}(x^*)|\partial_{x}[x^3 \pi(x,t)]/6\Omega\nonumber\\ &&+ \alpha_2(x^*)\partial_{x}^2\pi(x,t)/2.
		\label{eq:CritFP}
	\end{eqnarray}
Thus, $x(t)$ corresponds to diffusive motion in a quartic potential
	\begin{equation}
		V(x) \equiv  | \alpha_1^{\prime\prime\prime}(x^*)| x^4/24\Omega
	\end{equation} 
	with diffusion constant
	\begin{equation}
		D \equiv \alpha_2(x^*)/2~.
	\end{equation}
	This system is equivalently described by a self-adjoint Schrödinger-like operator \cite{risken}
	\begin{equation}
		L_x = D\partial_x^2 - V_S(x)
		\label{eq:CritSchrödinger}
	\end{equation}
	with 
	\begin{eqnarray}
		V_S(x) &\equiv& [V^{\prime}(x)]^2/4D - V^{\prime\prime}(x)/2\\&=& |\alpha_1^{\prime\prime\prime}(x^*)|^2x^6/72 \alpha_2(x^*)\Omega^2- |\alpha_1^{\prime\prime\prime}(x^*)|x^2/4\Omega~.\nonumber 
	\end{eqnarray} The eigenvalues $\lambda_n$ of this Schrödinger operator are real and nonnegative, i.e.,
	\begin{equation}
		L_x\psi_n(x) = - \lambda_n \psi_n(x)~, \lambda_n \geq 0.
	\end{equation}
	Moreover, they are equal to the eigenvalues of the Fokker-Planck operator defined through Eq. (\ref{eq:CritFP}). The corresponding eigenfunctions $\psi_n(x)$ determine the propagator of Eq. (\ref{eq:CritFP}) as 
	\begin{equation}
		\pi(x,t| x',t') = e^{[V(x') - V(x)]/2D}\sum_n\psi_n(x)\psi_n(x')e^{-\lambda_n(t-t')}~.
	\end{equation}
	This results in 
	\begin{equation}
		\int\limits_0^\infty dt\braket{x(t) x(0)} = \sum_n C_n/\lambda_n~,
		\label{eq:critGK}
	\end{equation}
	with 
	\begin{equation}
		C_n \equiv \int dx \int dx_0 ~xx_0  e^{[V(x_0) - V(x)]/2D}\psi_n(x)\psi_n(x_0) \pi^s(x_0)~,
		\label{eq:SchlöglCn}
	\end{equation}
	where $\pi^s(x_0)$ denotes the stationary solution of \req{eq:CritFP}.
	The $\Omega$-scaling of the correlation function, \req{eq:critGK}, is thus determined by the eigenvalues $\lambda_n$ and $C_n$. We find their dependencies on $\Omega$ in two steps. First, we examine the operator $L_x$. We rescale the $x$ variable according to
	\begin{equation}
		\tilde x \equiv \Omega^{-1/4} x~,
		\label{eq:SchloeglRescale}
	\end{equation}
	where a tilde means that the variable does not depend on the reaction volume $\Omega$. The Schrödinger operator transforms in this rescaled variable to
	\begin{equation}
		L_x = \tilde L_{\tilde x}/\sqrt{\Omega}~.
		\label{eq:SchlöglOperators}
	\end{equation}
	The operator $\tilde L_{\tilde x}$ is self-adjoint and independent of the system-size $\Omega$. Thus, its eigenvalues $\tilde \lambda_n$ and eigenfunctions $\tilde \Psi_n(\tilde x)$ are also independent of $\Omega$. These eigenvalues are connected to the ones of the Fokker-Planck operator as
	\begin{equation}
		\lambda_n = \tilde \lambda_n/ \sqrt\Omega~.
		\label{eq:SchloeglEigval}
	\end{equation}
	Furthermore, from the completeness relation, i.e.,
	\begin{equation}
		\delta_{n,m} = \int d\tilde x \int d\tilde x_0 \tilde \Psi_n(\tilde x)\tilde\Psi_m(\tilde x_0)~,
	\end{equation}
	we obtain the eigenfunctions of the original Schrödinger operator, Eq. (\ref{eq:CritSchrödinger}), as
	\begin{equation}
		\psi_n(x) \equiv \Omega^{-1/8} \tilde \Psi_n(\Omega^{-1/4}x)~.
	\end{equation}

		In the second step, we use these eigenfunctions to extract the scaling behavior of $C_n$, Eq. (\ref{eq:SchlöglCn}). Rescaling the integration variable according to Eq. (\ref{eq:SchloeglRescale}), we obtain
		
	\begin{widetext}
			\begin{eqnarray}
			C_n &=&\sqrt{\Omega} \int d\tilde x \int d\tilde x_0 ~\tilde x\tilde x_0  e^{[\tilde V(\tilde x_0) - \tilde V(\tilde x)]/2D}\tilde \Psi_n(\tilde x)\tilde \Psi_n(\tilde x_0) \tilde P(\tilde x_0) \equiv \sqrt{\Omega}~\tilde  C_n  ~.
			\label{eq:SchloeglConst}
		\end{eqnarray}
		With the scaling behavior of the eigenvalues, Eq. (\ref{eq:SchloeglEigval}), and the constant $C_n$, Eq. (\ref{eq:SchloeglConst}),  we obtain
		\begin{equation}
			( W_2^{+~\prime}[x^*]-W_2^{-~\prime}[x^*])^2 \int\limits_0^\infty dt\braket{x(t)x(0)}   =  \Omega~ ( W_2^{+~\prime}[x^*]-W_2^{-~\prime}[x^*])^2 \sum_n \tilde C_n/\tilde \lambda_n ~.
			\label{eq:SchloeglXX}
		\end{equation}
		Thus, we have derived that the first term of \req{eq:SchloeglDB} diverges linearly with the system-size $\Omega$ at the critical point.
		
We proceed by deriving a bound on the integral over $\braket{x(\tau)\xi_B(s)}$. We consider
		\begin{eqnarray}
			0 \leq \braket{\bigg( \int_0^t d\tau [c_xx(\tau)\pm c_\xi\xi_B(\tau)] \bigg)^2}& =& c_x^2 \int_0^t d\tau \int_0^t ds \braket{x(\tau)x(s)} + c_\xi^2\int_0^t d\tau \int_0^t ds \braket{\xi_B(\tau)\xi_B(s)} \nonumber\\&&\quad \pm 2 c_x c_\xi \int_0^t d\tau \int_0^t ds \braket{x(\tau)\xi_B(s)} ~.
		\end{eqnarray}
		with arbitrary constants $c_x$ and $c_\xi$. This leads to 
		\begin{equation}
			\pm c_x c_\xi \int_0^t d\tau \int_0^t ds \braket{x(\tau)\xi_B(s)}\leq  \frac 12 \bigg(c_x^2\int_0^t d\tau \int_0^t ds \braket{x(\tau)x(s)} +c_\xi^2 \int_0^t d\tau \int_0^t ds \braket{\xi_B(\tau)\xi_B(s)}\bigg)~.
			\label{eq:SchloeglMixed}
		\end{equation}
	Dividing both sides by $t$ and taking the limit $t \to \infty$, we obtain
		\begin{equation}
	\pm c_x c_\xi \lim\limits_{t\to\infty}\frac 1t\int_0^t d\tau \int_0^t ds \braket{x(\tau)\xi_B(s)}\leq  \frac 12 \bigg(c_x^2\int_0^\infty d\tau  \braket{x(\tau)x(0)} +c_\xi^2 \bigg)~.
	\label{eq:SchloeglMixed2}
\end{equation}
	\end{widetext}	

	Using \req{eq:SchloeglXX}, we find that the integral over the correlation function $\braket{x(\tau)\xi_B(s)}$ grows at most linearly in the system-size $\Omega$. Taking Eq. (\ref{eq:SchloeglXX}) and (\ref{eq:SchloeglMixed2}) into account, we, thus, have derived the divergence 
	\begin{equation}
		D^{\mathrm{crit}} \propto \Omega
		\label{eq:SchloeglDBCrit}
	\end{equation}
	for $\Omega \gg 1$. 
	
	We rationalize this result through the effect of critical slowing down. \cite{vankampen, gardiner} At the critical point, the inverse of the eigenvalues, \req{eq:SchloeglEigval}, and, hence, the time-scales for relaxation of fluctuations grow with the system-size $\Omega$ which drives the divergence of the diffusion coefficient $D$.
		
	We now determine the scaling of the diffusion coefficient as the critical point is approached.

	\subsubsection{Off the critical point}
	
	Moving away from the critical point, the macroscopic equation for the mean becomes
	\begin{equation}
		\partial_t \hat x(t) =  \alpha_1[\hat x(t)] = -[ \hat x(t) - { 1}/{\sqrt3}]^3 + \Delta  c_A [\hat x(t)^2 + 1/9]~.
		\label{eq:SchlöglCrit}
	\end{equation}
	This differential equation has a unique stable fixed point $x^* = x^*(\Delta  c_A)$ with $\alpha_1(x^*) = 0$ and $\alpha_1^{\prime}(x^*) < 0$, see Eq. (\ref{eq:FPabove}) and (\ref{eq:FPbelow}).  Thus,  the linear noise approximation leads to the following Langevin equations
		\begin{eqnarray}
			\partial_t x(t) &= &- |\alpha_1^{\prime}(x^*)|x(t) \nonumber\\&&+ \sqrt{c_Ax^*{}^2 + x^*{}^3}\xi_A(t) + 		 \sqrt{c_A/9+  x^*}\xi_B(t) ~,\nonumber\\
			\partial_t z_B(t)  &=& - x(t)+ 	\sqrt{c_A/9 + x^*}\xi_B(t) ~.
			\label{eq:critCLE}
		\end{eqnarray}
	For such a linear system, all integrals can be solved analytically, see Appendix \ref{app:linDif}. We thus obtain the diffusion coefficient
	\begin{eqnarray}
		D &=&{2^{1/3}}|\Delta c_A|^{-4/3}/{3^{5/6}} = \mathcal O(1),~~\text{for}~~\Omega \to \infty~,
		\label{eq:SchloeglDBNotCrit}
	\end{eqnarray}
	
	Comparing with the scaling ansatz, Eq. (\ref{eq:SchloeglScalingForm}), we obtain	$\varepsilon = -4/3$ and $\phi + \kappa\varepsilon=0~$. Furthermore, Eq. (\ref{eq:SchloeglDBCrit}) yields $\phi = 1$. Thus, we conclude
	\begin{equation}
		\phi = 1,~~~\kappa = 3/4,~~\text{and}~~\varepsilon = -4/3~,
	\end{equation}
	which analytically explains the numerical findings shown in \refFig{fig:2}.

\section{Along the pitchfork bifurcation}
\label{sec:pitchfork}
We turn to the pathway along the pitchfork bifurcation, i.e., the line which enters the cusp-like region tangential to its boundary, see the blue line in \refig{fig:pd}. It is parameterized as
\begin{equation}
	c_B(c_A) \equiv \sqrt 3/9 - (c_A - \sqrt 3)/3 =  \sqrt 3/9 - \Delta c_A/3~,
	\label{eq:BiLine}
\end{equation}
see Appendix \ref{app:pd} for a derivation of the slope. Thus, $\Delta c_A$ and $\Omega$ remain as free parameters of the reaction scheme \req{eq:SchloeglAB}.

\subsection{Numerical data}
\label{sec:BiNumerics}

We numerically integrate \req{eq:SchloeglAB} using the stochastic sampling algorithm \cite{gill77} while varying the control parameters. The particle flux $J$ and diffusion coefficient $D$ are determined according to Eqs. (\ref{eq:ZJB}) and (\ref{eq:ZDB}). 

\begin{figure}
		\centering
		\subfloat[\label{sfig:bi:2a}]{%
			\includegraphics[width=.4\textwidth]{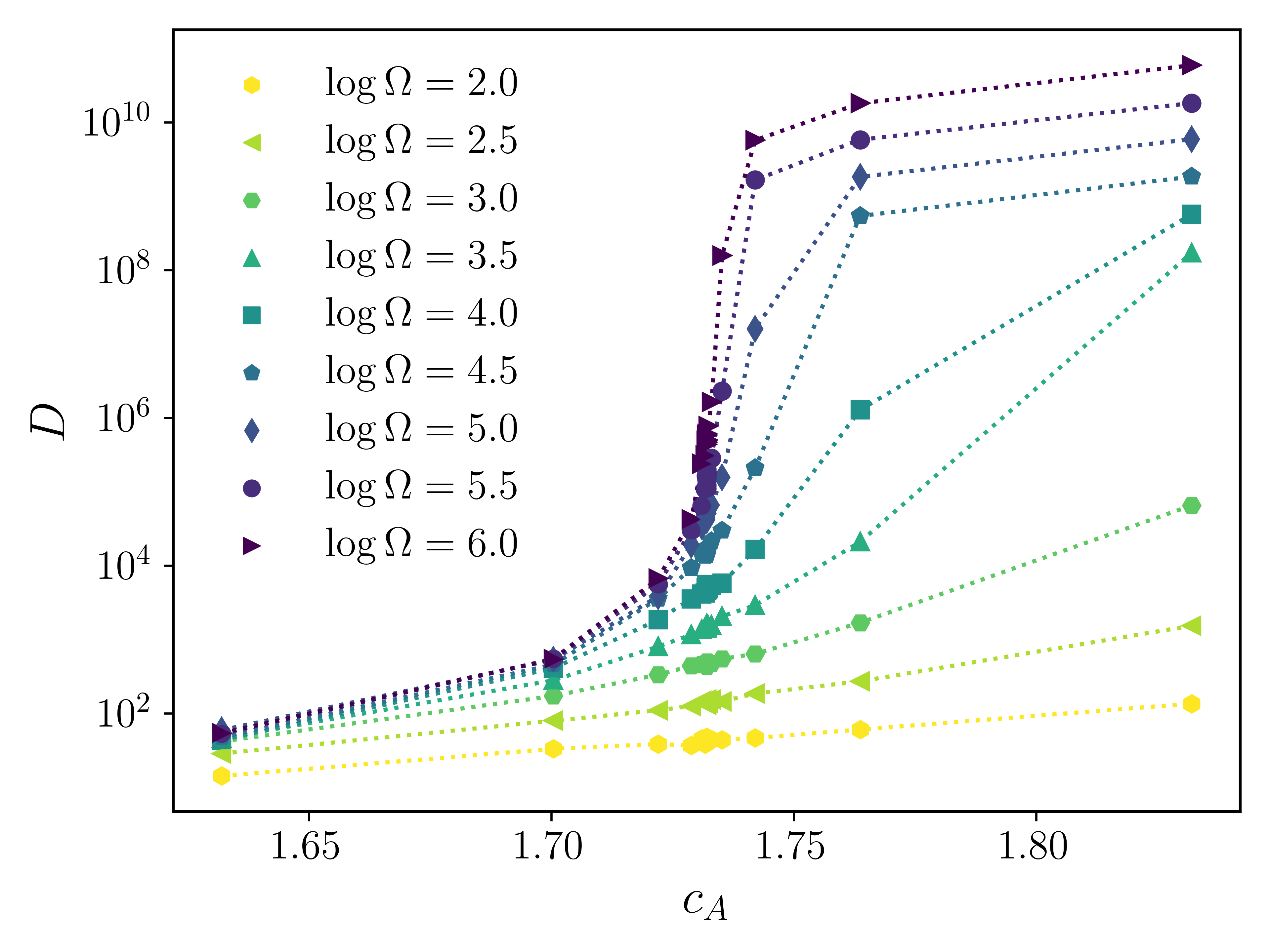}
		}	
		
		\subfloat[\label{sfig:bi:2b}]{%
			\includegraphics[width=.4\textwidth]{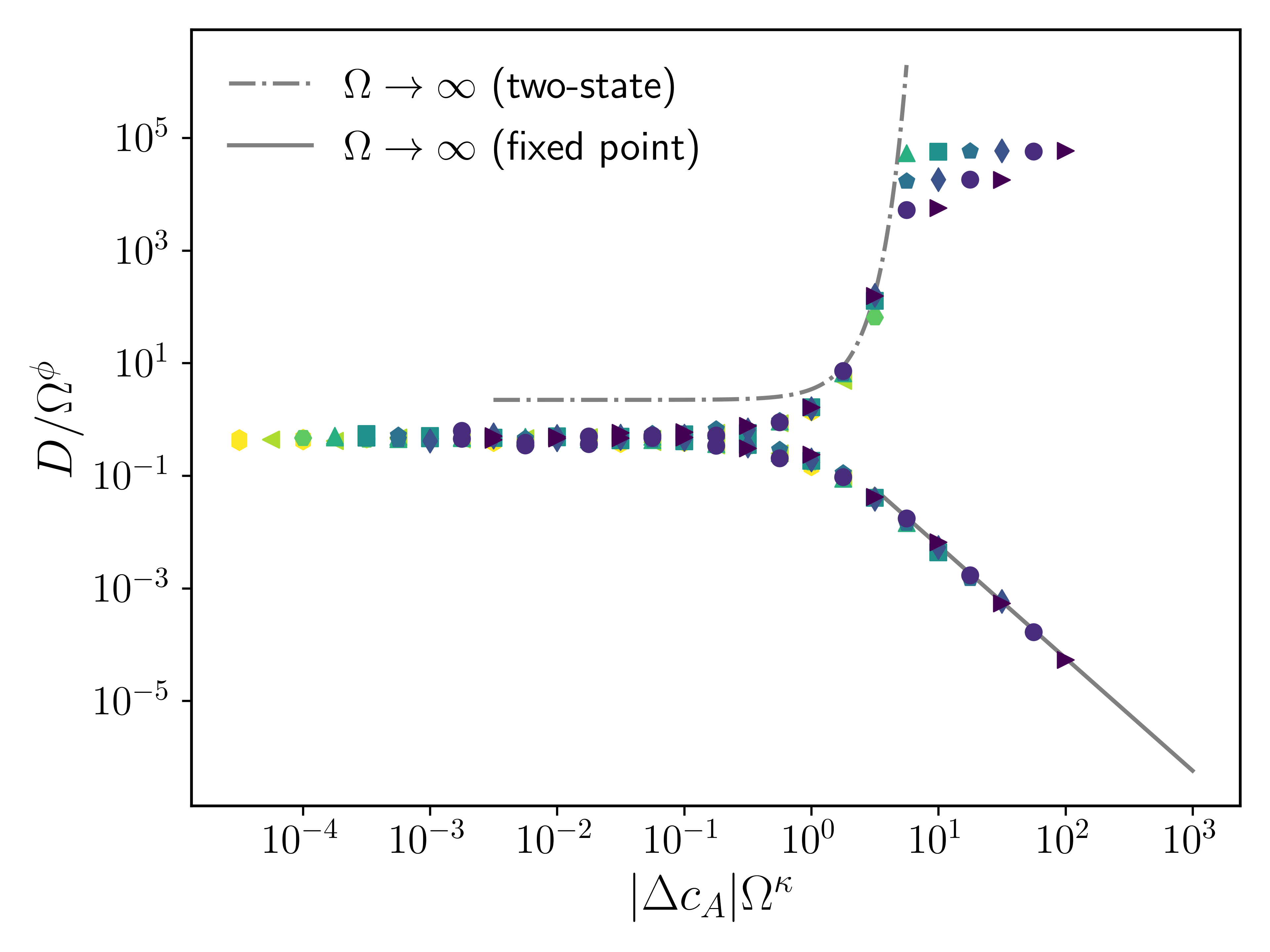}}
		
		\subfloat[\label{sfig:bi:2c}]{%
	\includegraphics[width=.4\textwidth]{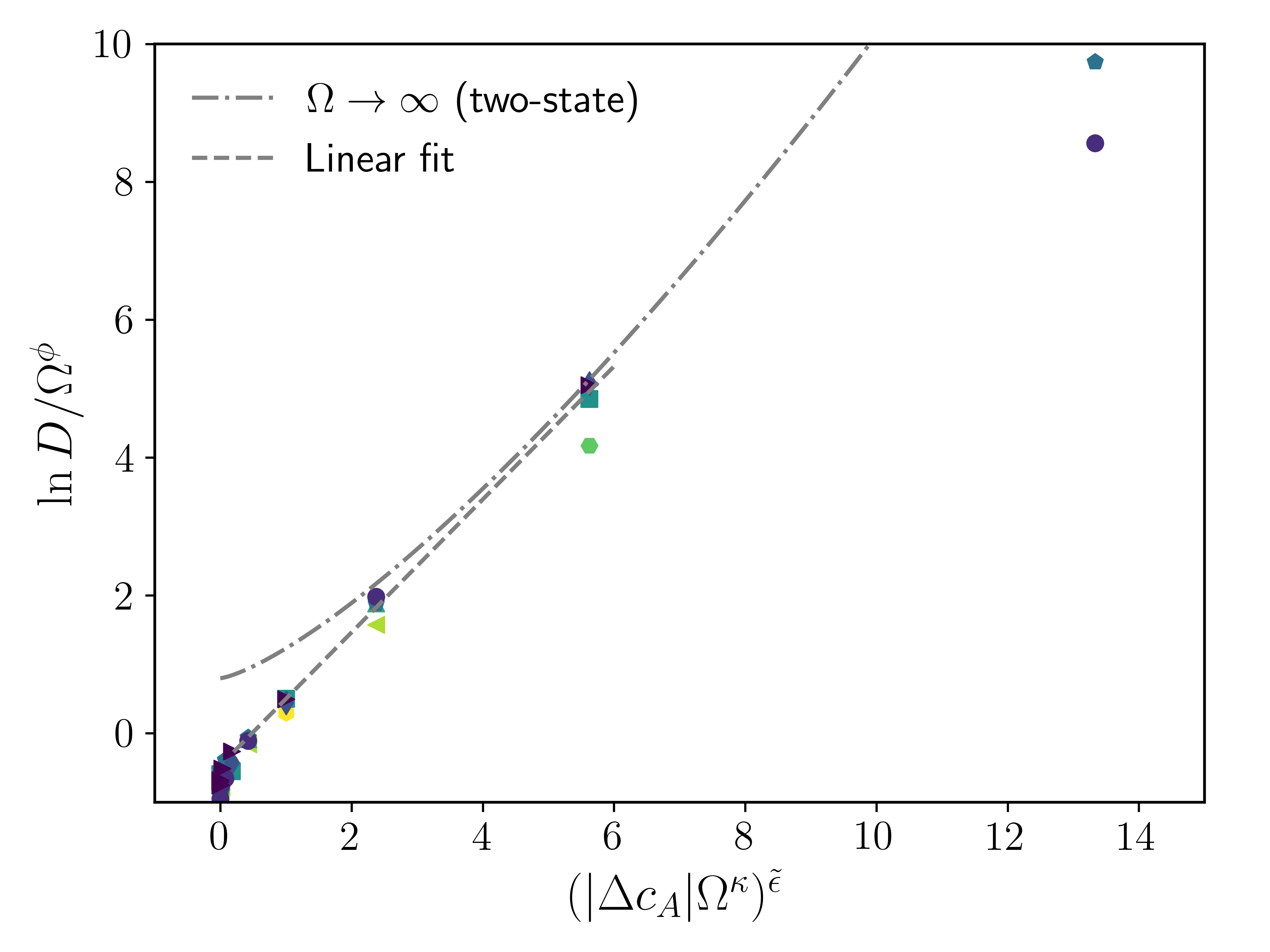}}		
	\caption{Diffusion coefficient $D$ of the Schlögl model at the pitchfork bifurcation. (a) $D$ as function of the control parameter $ c_A$ for various system sizes $\Omega$. (b) Scaling of $D$ according to $ D(\Delta  c_A, \Omega) = \Omega^\phi \mathcal D(|\Delta  c_A|~\Omega^\kappa)$ with $ c_A^{\mathrm{crit}} = \sqrt{3}$, $\kappa = 0.5$ and $\phi = 1.00$. (c) Logarithm of the re-scaled $D$ above the critical point as function of $(|\Delta c_A \Omega^\kappa|)^{\tilde \varepsilon}$ with $\tilde \varepsilon = 1.5$. The concentration $ c_B$ is chosen as function of $c_A$, Eq. (\ref{eq:BiLine}). The solid line represents the analytical result \req{eq:BiDBMono} and the dashed-dotted one \req{eq:BiDB}.}
		\label{fig:bi:2}
	\end{figure}

The diffusion coefficient increases with the concentration $c_A$ showing two distinct regimes, \refig{sfig:bi:2a}. Below the critical point, $D$ does not depend on the system size $\Omega$, whereas, it develops a dependence on $\Omega$ in the vicinity of $c_A^{\mathrm{crit}}$ and above. Rescaling the data according to 
\begin{equation}
	D = D(\Delta  c_A, \Omega) = \Omega^\phi \mathcal D(|\Delta  c_A|~\Omega^\kappa)~
\end{equation}
leads to the collapse of the data onto a master curve, see \refFig{sfig:bi:2b}. We determine the exponents as
\begin{equation}
	\phi = 1.0 \pm 0.1~~\text{and}~~\kappa = 0.50 \pm 0.05~.
	\label{eq:SchloeglBiNumericExp}
\end{equation}
This results in a scaling function 
\begin{equation}
	\mathcal D^<(x) \propto	\begin{cases}
		1~,& x \ll 1\\ x^\varepsilon~, & x \gg 1
	\end{cases}~
	\label{eq:SchloeglBiScalingForm}
\end{equation}
with 
\begin{equation}
	\varepsilon = - \phi/\kappa = -2.0\pm 0.2~,
	\label{eq:SchloeglBiNumericExp2}
\end{equation}
below the critical point. The plateau for small values of $|\Delta c_A|\Omega^\kappa$ indicates a scaling of the critical value as $D^{\mathrm{max}} \propto \Omega^\phi$ with $\phi$ given in Eq. (\ref{eq:SchloeglBiNumericExp}). The power-law behavior with $-\phi/\kappa$ below the critical point represents a scaling with $D \propto \Omega^0$ consistent with the raw data, see \refFig{sfig:bi:2a}.

In the supercritical regime, \refig{sfig:bi:2c}, we find
\begin{equation}
	\ln \mathcal D^>(x) \propto	\begin{cases}
		1~,& x \ll 1\\ x^{\tilde \epsilon}~, & x \gg 1
	\end{cases}~
	\label{eq:SchloeglBiScalingFormAbove}
\end{equation}
with
\begin{equation}
\tilde{\varepsilon }= 1.5 \pm 0.5~.
\end{equation}
 We attribute the outliers above the critical point to the fact that we leave the coexistence line in this regime, see \refig{fig:pd}. We only consider a tangential to the coexistence line and not the fully, curved line. Furthermore, the linear dependence between $(|\Delta c_A \Omega^\kappa|)^{\tilde \varepsilon}$ and $\ln D/\Omega^\phi$ observed in \refig{sfig:bi:2c} leads to our numerically obtained scaling function, \req{eq:SchloeglBiScalingFormAbove}.

In the next section, in order to obtain the scaling exponents analytically, we approximate the reaction scheme (\ref{eq:SchloeglAB}) below the critical point as in Sec. \ref{sec:SchloeglRate} and  \ref{sec:SchlöglTheory}. Above $c_A^{\mathrm{crit}}$, we can describe it as a two-state system.

		\begin{figure}
		\centering
		\subfloat[\label{sfig:bi:1a}]{%
			\includegraphics[width=.4\textwidth]{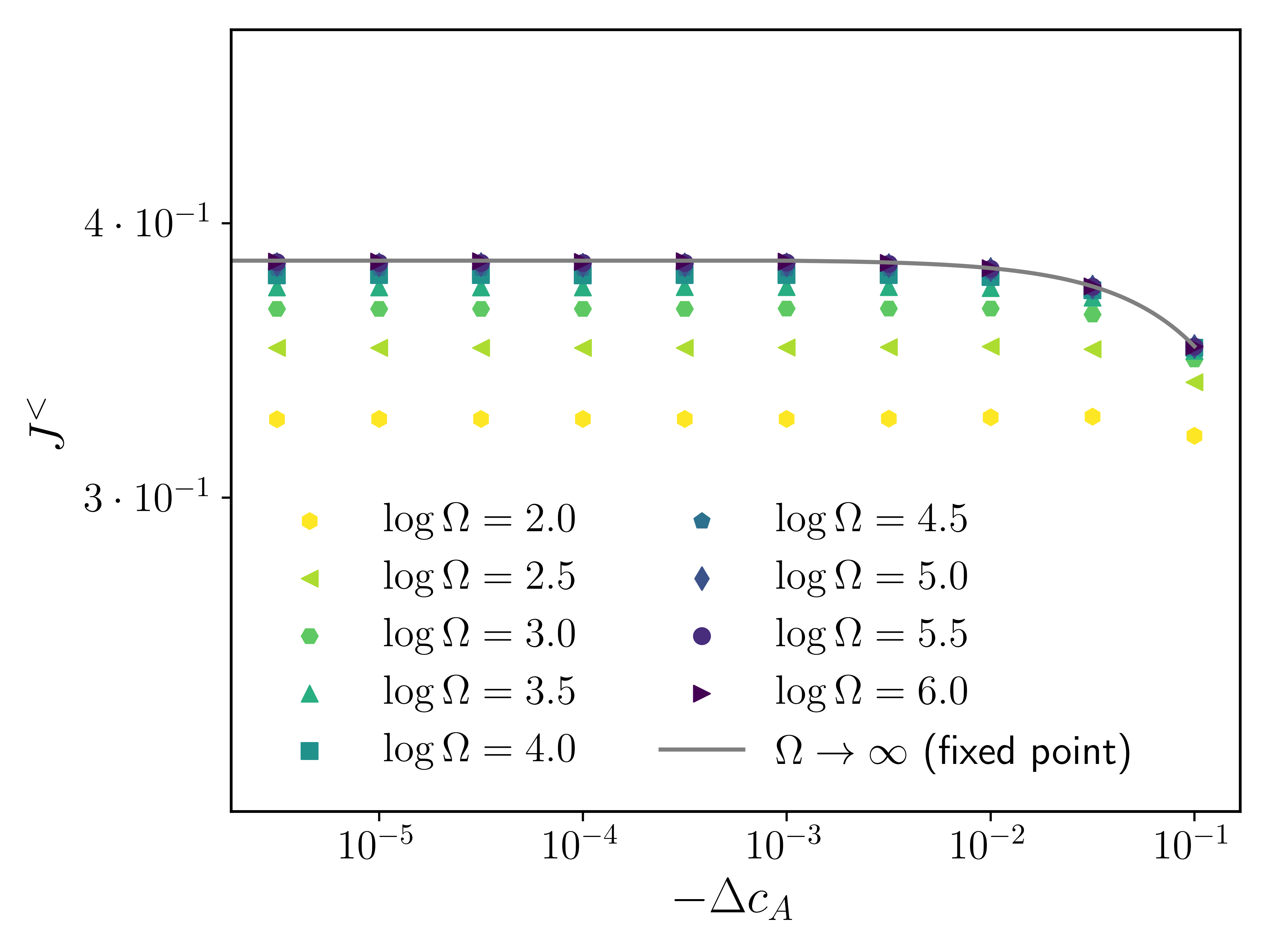}
		}	
		
		\subfloat[\label{sfig:bi:1b}]{%
			\includegraphics[width=.4\textwidth]{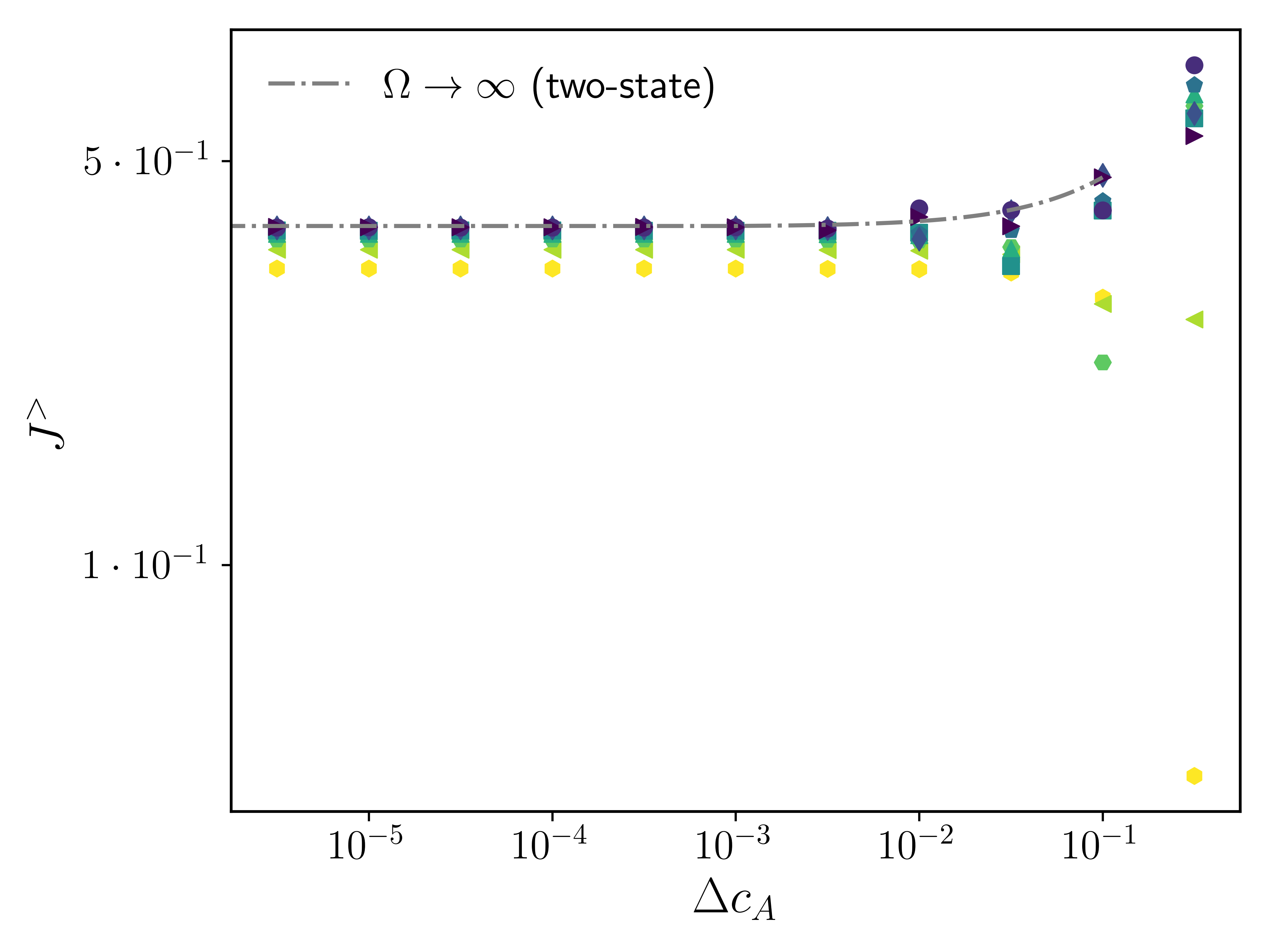}}
	\caption{Mean flux $J$ of the Schlögl model undergoing a pitchfork bifurcation for various system sizes $\Omega$ as function of the control parameter $ c_A$ (a) below and (b) above the critical point. The concentration $ c_B$ is determined according to Eq. (\ref{eq:BiLine}). The solid line represents \req{eq:BiJBMono} and the dashed dotted line \req{eq:BiJB}.}
		\label{fig:bi:1}
	\end{figure}

\subsection{In the mono stable regime and at the critical point}
\label{sec:BiMono}
The rate equation reads
\begin{equation}
	\partial_t\hat x(t) = - \hat x(t)^3 + \hat x(t)^2( \sqrt 3 + \Delta c_A) -\hat x(t) +  {\sqrt 3}/{9} - {\Delta c_A}/{3}~.
	\label{eq:BiRate}
\end{equation}
For $\Delta c_A < 0$, the only stable fixed point of this equation is
\begin{equation}
	x^* \equiv x_0 \equiv {1}/{\sqrt 3}~.
	\label{eq:x0}
\end{equation}
Thus, we find for the mean growth rate of particles $B$
\begin{equation}
	\partial_t \hat z_B(t) =x_0 -  ({\sqrt 3}/{9} - {\Delta c_A}/{3})  = {2 \sqrt 3 }/9+  {\Delta c_A}/{3} ~.
\end{equation}
The particle flux is hence given by
\begin{equation}
	J^< =  J^<(\Delta c_A) = {2 \sqrt 3 }/9+  {\Delta c_A}/{3}~.
	\label{eq:BiJBMono}
\end{equation}

In order to obtain the diffusion coefficient $D$, we employ the same technique as introduced in Sec. \ref{sec:SchlöglTheory}. Below the critical point, the chemical Langevin equation (\ref{eq:critCLE}) becomes
		\begin{eqnarray}
	\partial_t x(t) &= &- {2 {\sqrt 3}|\Delta c_A|x(t)}/3 \nonumber\\&+& \sqrt{{4\sqrt 3}/{9}+{\Delta c_A}/{3}}\xi_A(t)+ 		 \sqrt{{4}\sqrt 3/{9}-{\Delta c_A}/{3}}\xi_B(t) ~,\nonumber\\
	\partial_t z_B(t)  &=& - x(t)+ 	\sqrt{{4\sqrt 3}/{9}-{\Delta c_A}/{3}}\xi_B(t) ~.
	\label{eq:biCLE}
\end{eqnarray}
Solving all integrals (see Appendix \ref{app:linDif}), we obtain the diffusion coefficient in the mono stable regime
\begin{equation}
	D^< = 2\sqrt 3|\Delta c_A|^{-2}/3~.
	\label{eq:BiDBMono}
\end{equation}
At the critical point, we have the same set-up as in Sec. \ref{sec:SchlöglTheory}. Thus, we obtain the divergence 
\begin{equation}
	D^{<,\mathrm{crit}} \propto \Omega~.
\end{equation}

For the subcritical regime, we can summarize the scaling behavior of the diffusion coefficient $D$ in the scaling form
\begin{equation}
	D^{<} = \Omega^\phi \mathcal D^<(|\Delta c_A|\Omega^\kappa)
\end{equation}
with
\begin{equation}
	\phi = 1,~~\text{and}~~\kappa = 1/2~.
	\label{eq:bi:exp1}
\end{equation}
The scaling form is
\begin{equation}
	\mathcal D^<(x) \propto	\begin{cases}
		1~,& x\ll 1\\ x^\varepsilon~,  & x \gg 1
	\end{cases}~
\end{equation}
with exponent
\begin{equation}
\varepsilon = - \phi/\kappa = -2~.	
\end{equation}
Thus, approaching the critical point from below, we analytically find the same scaling exponents as obtained numerically.

\subsection{Bistable regime and two-state approximation}
\label{sec:BiBi}
We consider \req{eq:BiLine} for $\Delta c_A >0$, i.e., in the bistable regime. Here, the steady state dynamics is given by two stable fixed points,
\begin{eqnarray}
x_\pm(\Delta c_A) &\equiv & ( 2 \sqrt 3 + 3 \Delta c_A \pm [24\sqrt 3 \Delta c_A +  9\Delta c_A^2]^{1/2})/6 \nonumber \\ &\approx &1/\sqrt 3 \pm \sqrt{2\Delta c_A}/3~,
\end{eqnarray}
that are separated by an unstable fixed point $x_0$, \req{eq:x0}. The system randomly jumps between these two steady states. In the following, we treat the system as a two-state system in order to obtain an estimate for the particle flux $J$ and diffusion coefficient $D$. \cite{hang84,vell09,nguy20}

In the limit of large system-sizes, $\Omega \gg 1$, the steady state solution of reaction scheme (\ref{eq:SchloeglCRN}) is given by
\begin{equation}
	p^s(x) = \mathcal N \exp[ - \Omega \phi(x)]~,
\end{equation}
where we introduced the nonequilibrium potential
\begin{equation}
	\phi(x) \equiv - \int_0^x du \ln \frac{c_A u^2 + c_B}{ u^3 +  u}~.
	\label{eq:noneqPot}
\end{equation}
The fixed points of \req{eq:BiRate} determine the extrema of this potential with a maximum at $x_0$ and minima at $x_\pm$.

The transition rates $r_\pm$ from one minimum to the other are given by the potential barrier between the minima and their curvature as\cite{hang84,vell09}
\begin{equation}
r_\pm \equiv r_{x_\pm \to x_\mp} \equiv \frac{F(x_\pm)\sqrt{-\phi^{\prime \prime}(x_0)\phi^{\prime \prime}(x_\pm)}}{2\pi\Omega} e^{-\Omega[\phi(x_0) - \phi(x_\pm)]}~,
\label{eq:BiRates}
\end{equation}
where a prime denotes a derivative with respect to the argument. We have introduced $F(x) \equiv (\sqrt 3 + \Delta c_A)  x^2 + \sqrt{3}/9 - \Delta c_A/3$ for briefness of notation.
Furthermore, along the coexistence line, the minima are equivalent, i.e.,
\begin{equation}
	\phi(x_+) = \phi(x_-)~~~\text{and}~~~\phi^{\prime \prime}(x_+) = \phi^{\prime \prime}(x_-)~.
\end{equation}
The probability $p_\pm$ for the system to be in either of the minima is
\begin{equation}
	p_\pm \equiv r_\mp/(r_++r_-) = 1/2~,
\end{equation}
due to the symmetry of the problem.

Following Ref. \cite{nguy20}, the particle flux is in leading order determined by
\begin{eqnarray}
	J^> &\equiv& p_- J^- + p_+ J^+ \nonumber\\
	 & \approx &  {2}/{3\sqrt 3} + {5}\Delta c_A/{6}~,
	 \label{eq:BiJB}
\end{eqnarray}
where $J^\pm$ denotes the mean flux of species $B$ caused through the respective minimum, i.e.,
\begin{equation}
	J^\pm \equiv  x_\pm(\Delta c_A)-c_B(\Delta c_A)~.
\end{equation}
The diffusion coefficient of the two-state model is given by
\begin{eqnarray}
D^> &\equiv& {p_+p_-}(J^+ - J^-)^2/({r_++r_-}) +\mathcal O(\Omega)\nonumber \\
& = & \Omega {\pi} \exp({\sqrt 3}\Delta c_A^2 \Omega/{4}) /{\sqrt 2} +\mathcal O(\Omega)~\label{eq:BiDB}\\
 & \equiv& \Omega \mathcal D^>(|\Delta c_A|\Omega^{\frac 12})\nonumber~.
\end{eqnarray}
In the limit $\Delta c_A \to 0$, this expression leads to
\begin{equation}
	D^{>,\mathrm{crit}}\equiv D|_{\Delta c_A = 0} = \mathcal O(\Omega)~,
\end{equation}
which is consistent with the result derived in the previous section, i.e., we obtain the scaling exponent
\begin{equation}
	\phi = 1~.
	\label{eq:bi:exp2}
\end{equation}
Furthermore, $D^>/\Omega$ is a function of $|\Delta c_A|\Omega^{1/2}$, \req{eq:BiDB}. Thus, we get the scaling exponent
\begin{equation}
	\kappa =  1/2~,
	\label{eq:bi:exp3}
\end{equation}
which agrees with the one below the critical point. However, in the supercritical regime, we find an exponential behavior, \req{eq:BiDB}, with exponent
\begin{equation}
\tilde \varepsilon = 2~,
\label{eq:bi:exp4}
\end{equation} 
which is consistent with the numerically obtained one, \req{eq:SchloeglBiScalingFormAbove}.

In summary, we have analytically derived the particle flux $J^>$, \req{eq:BiJB}, which continuously extends the one obtained in the monostable regime $J^<$, \req{eq:BiJBMono}, through the continuous phase transition following the tangential line into the bistable region, \req{eq:BiLine}. Furthermore, approximating the system as a two-state one, we have obtained a theoretical prediction for the scaling of $D$, \req{eq:BiDB}.

\section{Scaling form}
\label{sec:scalingform}
\subsection{General monostable pathway}
\label{sec:delta}
In order to obtain a general scaling form for the diffusion coefficient $D$, we consider a genuine line in the $c_A-c_B$-plane through the critical point. For calculations it is convenient to use the following parameterization
\begin{equation}
	c_B(\Delta c_A) \equiv {\sqrt 3}/{9} - (1/3 - \delta)\Delta c_A~.
	\label{eq:DeltaLine}
\end{equation}
In the limit $\delta \to 0$, we recover the line along the pitchfork bifurcation, \req{eq:BiLine}. $\Delta c_A$ and $\delta$ remain as control parameters.

The rate equation reads
\begin{equation}
	\partial_t\hat x(t) = - \hat x(t)^3 + \hat x(t)^2( \sqrt 3 + \Delta c_A) -\hat x(t) + {\sqrt 3}/{9} - (1/3 - \delta)\Delta c_A~.
	\label{eq:DeltaRate}
\end{equation}
For $\delta \neq 0$, in the limit of $|\Delta c_A|\ll 1$, the fixed point of this equation is given by
\begin{equation}
	x^*(\Delta c_A,\delta) \equiv {1}/{\sqrt 3} + \text{sign}(\Delta c_A\delta)|\Delta c_A\delta|^{ 1/3}~.
	\label{eq:DeltaFix}
\end{equation}
Following the techniques introduced in Sec. \ref{sec:SchloeglRate}, we obtain
\begin{eqnarray}
	J(\Delta c_A,\delta) &\equiv&  {2\sqrt 3}/{9}+ \text{sign}(\Delta c_A\delta)|\Delta c_A\delta|^{1/3} + \mathcal O(\Delta c_A)\nonumber \\ & \propto& |\Delta c_A \delta|^{1/3}~.
	\label{eq:deltajb}
\end{eqnarray}
Thus, we find a continuous particle flux as function of $\Delta c_A\delta$.

Off the critical point, using the linear noise approximation introduced in Sec. \ref{sec:SchlöglTheory}, we obtain for the diffusion coefficient
\begin{equation}
	D(\Delta c_A, \delta) = {8\sqrt3} |\Delta c_A\delta|^{-4/3}/{27}~.
	\label{eq:DeltaD}
\end{equation}
Right at the critical point, the diffusion coefficient is given by \req{eq:SchloeglDBCrit}.

In summary, for any $\delta \neq 0$, we obtain the same scaling behavior for the particle flux and diffusion coefficient as along the critical line.

\subsection{Change of coordinates}
\label{sec:coordinates}
\begin{figure}
	\centering
	\includegraphics[width=0.4\textwidth]{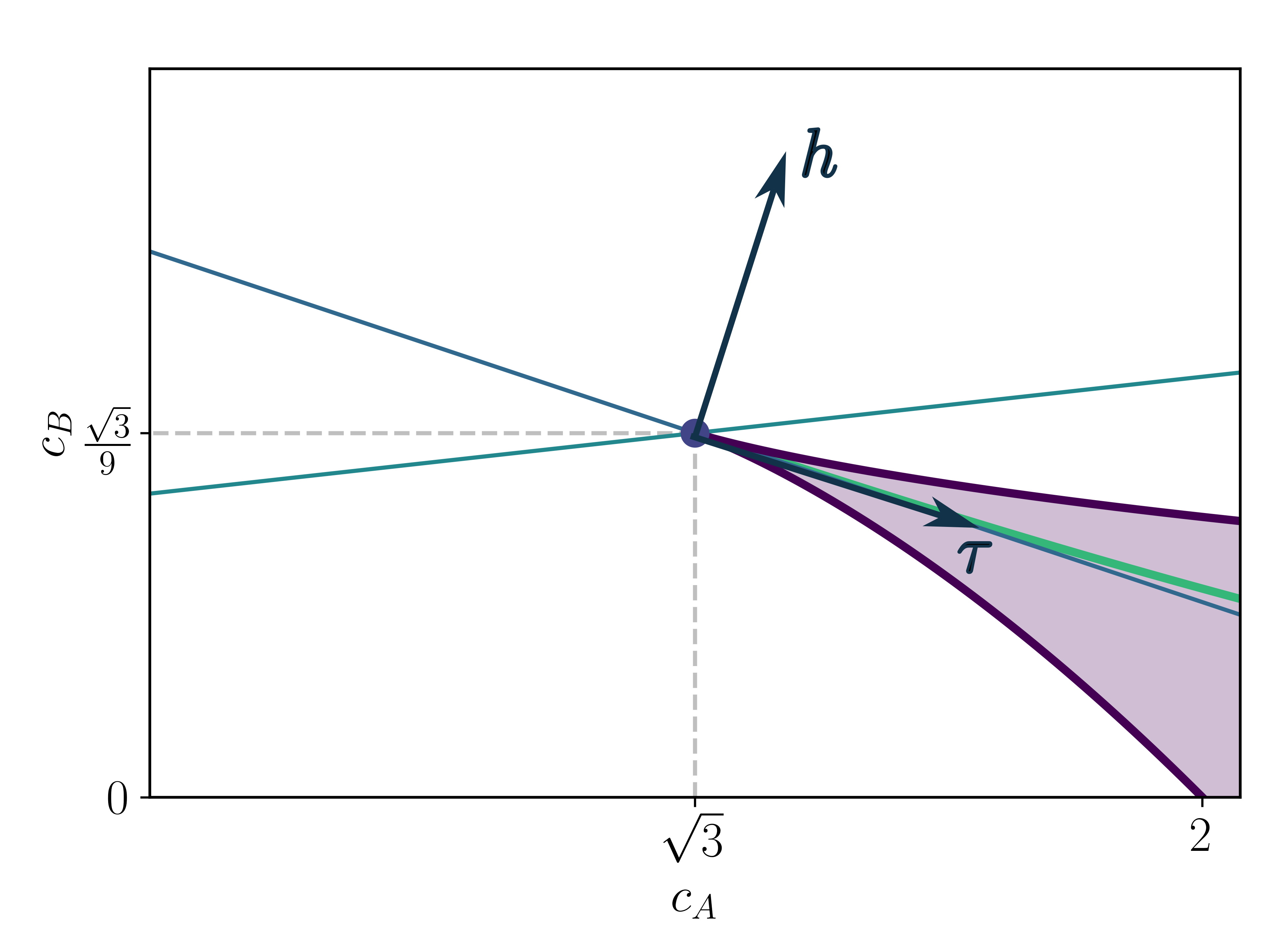}
	\caption{Phase-diagram of the Schlögl model, \req{eq:SchloeglCRN}, with the coordinate-frame for the new variables $(\tau, h)$.}
	\label{fig:coordinates}
\end{figure}

We change from the $(c_A, c_B)$-parameterization into a frame with the origin at the critical point. Furthermore, we consider the basis with direction parallel to the line along the pitchfork bifurcation and perpendicular to it, see \refig{fig:coordinates}. 

We denote the deviations from the critical point along the $c_A$-axis as $\Delta c_A$ and along the $c_B$-axis as $\Delta c_B \equiv c_B - c_B^{\mathrm{crit}}$. In the direction along the pitchfork bifurcation, we label the deviations $\tau$ and into the perpendicular direction $h$, see \refig{fig:coordinates}. The coordinate transformation is given by
\begin{equation}
	\begin{pmatrix} \Delta c_A \\ \Delta c_B \end{pmatrix} = \begin{pmatrix} 3/\sqrt{10} & 1/\sqrt{10} \\ -1/\sqrt{10} & 3/\sqrt{10}\end{pmatrix}~\begin{pmatrix}\tau \\ h \end{pmatrix}~.
\end{equation}

For the line along the pitchfork bifurcation, \req{eq:BiLine}, we obtain the parameterization
\begin{equation}
	\tau = \sqrt{10}\Delta c_A/3,~~\text{and}~~h = 0~.
	\label{eq:paraBi}
\end{equation}
For any other line through the critical point, \req{eq:DeltaLine}, we find
\begin{equation}
	\Delta c_A = (3\tau +h)/\sqrt 10 ~~\text{and}~~\Delta c_B = 10h/(9\tau +3h)~.
	\label{eq:paraCrit}
\end{equation}
Thus, we get
\begin{equation}
	\Delta c_A \delta = \sqrt{10}h/3~,
\end{equation}
which is independent of $\tau$.

\subsection{Scaling form for the particle flux}
With the parameterization introduced in the previous section, \req{eq:paraBi} and (\ref{eq:paraCrit}), and the results for the particle fluxes, \req{eq:BiJBMono}, (\ref{eq:BiJB}) and (\ref{eq:deltajb}), we find the following scaling form for the particle flux in the vicinity of the critical point
\begin{equation}
	J(\tau,h) - J(0,0) =  \begin{cases} \tau/\sqrt{10}~,&\tau < 0,~h=0\\ 5\sqrt{10}\tau/20~,&\tau >0,~h=0\\ 10^{1/6}\text{sign}(h)|h|^{1/3}/3^{1/3}~,& h\neq 0\end{cases}~,
	\label{eq:scalingJ}
\end{equation}
with $J(0,0) \equiv 2\sqrt 3/9$. For non-vanishing $h$, the derivative with respect to $h$ scales as
\begin{equation}
	\frac{\partial J(\tau,h)}{\partial h}\propto |h|^{-2/3}~,
\end{equation}
which agrees with the result obtained along the critical line, i.e., we recover the scaling exponent, \req{eq:chi}. 

\subsection{Scaling form for the diffusion coefficient}
Using the results of Sec. \ref{sec:coordinates} and \ref{sec:pitchfork}, we find the following scaling form for the diffusion coefficient
\begin{equation}
	D = D(\Omega, \tau, h) = \Omega^\phi \mathcal D(\Omega^{\kappa_\tau} \tau, \Omega^{\kappa_h}h)~.
	\label{eq:scalingform}
\end{equation}
In the special case of $h = 0$, we have
\begin{equation}
	D(\Omega, \tau, 0) = \Omega^\phi \mathcal D(\Omega^{\kappa_\tau}\tau,0)~.
\end{equation}
The scaling function is
\begin{equation}
	\mathcal D(x,0) \propto \begin{cases} 1~,&|x| \ll 1 \\ (-x)^{\epsilon_\tau^<}~,&x \ll-1 \\ \exp(4x^{\epsilon_\tau^>}/3)~,& x\gg 1 \end{cases}~,
	\label{eq:critForm1}
\end{equation}
with scaling exponents
\begin{equation}
	\phi = 1,~~\kappa_\tau = 1/2,~~\text{and}~~\epsilon_\tau^> = -\epsilon_\tau^< = 2~.
	\label{eq:crit1}
\end{equation}

For $h \neq 0$ and any $\tau$, we obtain
\begin{equation}
	\mathcal D(x,y) \propto \begin{cases} 1~,&0<|y|\ll 1 \\ |y|^{\epsilon_h}~,& |y|\gg 1\end{cases}~,\label{eq:critForm2}
\end{equation}
with critical exponents
\begin{equation}
	\kappa_h = 3/4~~\text{and}~~\epsilon_h = -4/3~,
	\label{eq:crit2}
\end{equation}
compare Sec. \ref{sec:SchlöglTheory} and \ref{sec:delta}.

In summary, using the new coordinates $(\tau,h)$, we have derived a scaling form for the diffusion coefficient $D$ in the vicinity of the critical point, \req{eq:scalingform}. The scaling functions for the two different regimes are given in \req{eq:critForm1} and (\ref{eq:critForm2}) and the critical exponents in \req{eq:crit1} and (\ref{eq:crit2}). 

\section{Discussion and conclusion}
\label{sec:Conclusion}

We have numerically as well as analytically derived the scaling exponents of the thermodynamic flux and its associated diffusion coefficient for the univariate Schlögl model undergoing a continuous phase transitions. As representative path in phase space, we have considered the critical line, \req{eq:CritLine}, along which the system exhibits critical scaling. Based on the rate-equation, we have obtained the mean behavior of the particle flux, Sec. \ref{sec:SchloeglRate}, which is in good agreement with the numerical found behavior, Sec. \ref{sec:SchlöglNumerics}. Using the chemical Langevin equation and van Kampen's system-size expansion, we have analytically determined the critical exponents of the diffusion coefficient associated with the flux of the species in the reservoir. 

For the second path, we have numerically shown that the system undergoes a continuous transition, Sec. \ref{sec:BiNumerics}. The fluctuations characterized through the associated diffusion coefficient diverge linearly in the system-size at the critical point. Furthermore, the diffusion coefficient obeys a scaling form, thus, indicating critical behavior. Using the techniques introduced in Sec. \ref{sec:SchloeglRate} and \ref{sec:SchlöglTheory}, we have obtained the scaling exponents in the monostable, subcritical regime, Sec. \ref{sec:BiMono}. Applying a two-state approximation, we have calculated the mean behavior of the particle flux and its diffusion coefficient in Sec. \ref{sec:BiBi}. We have obtained critical exponents that agree with the one of the subcritical regime. Furthermore, we have derived an exponential scaling function above the critical point which is consistent with previous results.\cite{nguy20} 

In Sec. \ref{sec:scalingform}, we have transformed to the variables parallel to the line along the pitchfork bifurcation and the direction perpendicular to it. Parameterizing the $(c_A,c_B)$-plane in those coordinates, we have derived a scaling form for the particle flux, \req{eq:scalingJ}, and the diffusion coefficient, \req{eq:scalingform}, that summarizes the results of the previous sections.

In conclusion, we have examined the thermodynamic flux and its fluctuations in the vicinity of the critical point of the Schlögl model. We have analytically determined critical exponents based on the chemical Langevin equation, van Kampen's system size expansion and a two-state approximation. Furthermore, this paper complements previous studies that have  considered the first order transition when crossing the coexistence line in \refig{fig:pd} \cite{nguy20,ge09}. Thus, our contribution completes the understanding of the thermodynamic treatment of the reaction scheme (\ref{eq:SchloeglCRN}).

The critical behavior as derived here for the Schlögl model should be universal for any nonequilibrium reaction network with a critical point where a first-order transition between two stable dynamical phases terminates. A different universality class can be expected for more complex reaction schemes like the Brusselator \cite{prig68,schn79,lefe88,qian02c,nguy18} where a Hopf bifurcation separates an oscillatory regime described by a limit cycle from a monostable regime. It will be interesting to apply the concepts derived here to this complementary paradigm of a nonequilibrium phase transition.

\appendix

\section{Phase diagram in the $c_A-c_B$-plane}
\label{app:pd}
We derive the phase diagram,  \refig{fig:pd}. The right hand side of the rate-equation (\ref{eq:SchloeglRateEquation}) is a cubic polynomial. The number of roots, i.e., fixed points, is determined by the cubic discriminant
\begin{equation}
	\Delta \equiv - 4 + c_A^2 + 18 c_Ac_B - 4 c_A^3c_B - 27 c_B^2~.
\end{equation}
For $\Delta > 0$, the polynomial has three distinct real roots. Whereas, for $\Delta < 0$ there is only one real root. The limiting case of $\Delta = 0$ gives the phase boundary where we find at least one double root. Thus, setting $\Delta = 0$ and solving for $c_B$, we obtain the boundary curves of the cusp-like region
\begin{equation}
	c^\pm_B(c_A) \equiv [9 c_A - 2 c_A^3 \pm (c_A^2 - 3)^{3/2}]/27~.
\end{equation}
At the critical point, both solution meet, i.e.,
\begin{equation}
	c^\pm_B(c_A \to c_A^{\mathrm{crit}}) = {\sqrt{3}}/{9}~,
\end{equation}
with the same slope
\begin{equation}
	\frac{\partial c_B^\pm}{\partial{c_A}}(c_A^{\mathrm{crit}}) = -  1/3~.
\end{equation}
This is the slope of the blue line, \refig{fig:pd}, considered in Sec. \ref{sec:pitchfork}.

In the regime with two stable fixed points, $x_\pm$, see Sec. \ref{sec:pitchfork}, there exists a line in phase space which is denoted as coexistence line. This line is defined through the condition that the nonequilibrium potential, \req{eq:noneqPot}, has two global minima, i.e.,
\begin{equation}
	\phi(x_+) = \phi(x_-)~.
\end{equation}
Since the potential and the two minima depend on the concentrations $c_A$ and $c_B$, above relation defines the coexistence line as a parametric curve in the $c_A-c_B$-plane.

\section{Diffusion coefficient in linear system}
\label{app:linDif}

Consider a system that is described by the following Langevin equation
\begin{eqnarray}
	\partial_t x(t) &=&  - K x(t) + \sqrt{2\Delta_A} \xi_A(t) + \sqrt{2\Delta_B}\xi_B(t)\nonumber\\
	\partial_t z(t) &=& - x(t) + \sqrt{2\Delta_B}\xi_B(t)~,
\end{eqnarray}
with $K,\Delta_A,\Delta_B>0$ and Gaussian white noise $\xi_i(t)$, i.e., $\braket{\xi_i(t)} = 0$ and $\braket{\xi_i(t)\xi_j(t^\prime)} = \delta_{ij}\delta(t-t^\prime)$.
The diffusion coefficient,
\begin{equation}
	D \equiv \lim\limits_{t\to\infty} \frac{\text{Var}[z(t)]}{2 t}~,
\end{equation}
can be obtained explicitly as follows. 

The $x$-fluctuations describe an Ornstein-Uhlenbeck process\cite{gardiner,risken} with  solution
	\begin{equation}
	x(t) = \exp(-K t) \int\limits_0^td \tau \exp(K \tau )\sum\limits_{\rho = A,B}\sqrt{2\Delta_\rho}\xi_\rho(\tau )~.
\end{equation}
A solution for $z(t)$ is given by
\begin{equation}
	z(t) = - \int\limits_0^t d\tau x(\tau) + \sqrt{2\Delta_B}\int\limits_0^t d\tau \xi_B(\tau)~.
\end{equation}
With 
\begin{equation}
	\braket{x(\tau)x(\tau^\prime)} = (\Delta_A+\Delta_B)( e^{-K|\tau -\tau^\prime|} - e^{-K(\tau + \tau^\prime)})/{K} 
\end{equation}
and
\begin{equation}
	\braket{x(\tau)\xi_B(\tau^\prime)} = \sqrt{2\Delta_B} \exp[-K(\tau-\tau^\prime)]\theta(\tau-\tau^\prime)~,
\end{equation}
where $\theta(x) = 1$ for $x>0$ and $\theta(x) =0$ elsewhere, we get
\begin{equation}
	D = \Delta_B (1 - 1/K)^2+\Delta_A/{K^2}~.
\end{equation}
Identifying the coefficients as
\begin{eqnarray}
	K = |\alpha_1^\prime(x^*)|,~\Delta_A = (\sqrt{3} +\Delta c_A) x^*{}^2+x^*{}^3\\~~~~~~~~~~\text{and}~~\Delta_B = (\sqrt 3 + \Delta c_A)/9+x^*~\nonumber
\end{eqnarray}
where $x^* = x^*(\Delta c_A)$ is given in Eq. (\ref{eq:SchloeglFP}) and $\alpha_1(x)$ in Eq. (\ref{eq:SchlöglCrit}), we obtain Eq. (\ref{eq:SchloeglDBNotCrit}) after a Taylor-expansion for $|\Delta c_A|\ll 1$.

Respectively, identifying the coefficients as
\begin{eqnarray}
	K = |\alpha_1^\prime(x_0)|,~\Delta_A = (\sqrt{3} +\Delta c_A) x_0^2+x_0^3\\~~~~~~~~~~\text{and}~~\Delta_B = (\sqrt 3/9 - \Delta c_A/3)+x_0~\nonumber
\end{eqnarray}
where $x_0 = 1/\sqrt 3$ is given in Eq. (\ref{eq:x0}) and $\alpha_1(x)$ through the right hand side of the rate-equation (\ref{eq:BiRate}), we obtain Eq. (\ref{eq:BiDBMono}).

The following coefficients
\begin{eqnarray}
	K = |\alpha_1^\prime(x^*)|,~\Delta_A = (\sqrt{3} +\Delta c_A) x^*{}^2+x^*{}^3\\~~~~~~~~~~\text{and}~~\Delta_B = \sqrt 3/9 - (1/3 -\delta)\Delta c_A +x^*~\nonumber~,
\end{eqnarray}
where $x^* = x^*(\Delta c_A\delta)$ is given in Eq. (\ref{eq:DeltaFix}) and $\alpha_1(x)$ through the right hand side of Eq. (\ref{eq:DeltaRate}), lead to Eq. (\ref{eq:DeltaD}).

		\bibliography{../../my-awesome-bib/refsDiss} 

	\end{document}